\documentclass{aa}
\usepackage{graphicx}
\usepackage{multirow}
\usepackage{natbib}
\usepackage{ulem}
\usepackage{txfonts}
\usepackage{hyperref}
\hypersetup{colorlinks=true, linkcolor=red, citecolor=blue, urlcolor=blue}

\def\void#1{{}}

\newcommand{\Ks}{K_s}
\newcommand{\ab}{}

\newcommand{\txn}[1]{\textnormal{#1}}

\newcommand{\Msun}{\hbox{$\txn{M}_{\odot}$}}

\newcommand*\lephare{L\textsc{e} P\textsc{hare}~}

\usepackage{color}

\begin{document}

\title{The VIPERS Multi-Lambda Survey. I} 
\subtitle{ UV and NIR observations, multi-colour catalogues, and photometric redshifts} 
\author{T.~ Moutard\inst{1}  \thanks{thibaud.moutard@lam.fr}
\and S. ~Arnouts \inst{1}
\and O. ~Ilbert\inst{1}
\and J. ~Coupon\inst{2}
\and P. ~Hudelot\inst{3}
\and D. ~Vibert\inst{1}
\and V. ~Comte\inst{1}
\and S. ~Conseil\inst{1}
\and \\I. ~Davidzon\inst{1,4}
\and L.~Guzzo\inst{5}
\and A. ~Llebaria\inst{1}
\and C. ~Martin\inst{6}
\and H. ~J. ~McCracken\inst{3}
\and B. ~Milliard\inst{1}
\and G. ~Morrison\inst{7,8}
\and \\D. ~Schiminovich\inst{9}
\and M. ~Treyer\inst{1}
\and L. ~Van ~Werbaeke\inst{10} 
}
\institute{
Aix Marseille Universit\'e, CNRS, LAM - Laboratoire d'Astrophysique de Marseille, 38 rue F. Joliot-Curie, F-13388, Marseille, France 
\and
Astronomical Observatory of the University of Geneva, ch. d Ecogia16, 1290 Versoix, Switzerland
\and
CNRS, UMR7095 \& UPMC  Institut d'Astrophysique de Paris, 98 bis Boulevard Arago, 75014 Paris, France
\and
INAF - Osservatorio Astronomico di Bologna, via Ranzani 1, I-40127, Bologna, Italy
\and
INAF - Osservatorio Astronomico di Brera, via E. Bianchi 46, 23807 Merate/via Brera 28, 20122, Milano, Italy
\and
California Institute of Technology, Pasadena, CA 91125, USA
\and
Institue for Astronomy, University of Hawaii, Honolulu, HI, 96822, USA
\and
Canada-France-Hawaii Telescope, 65-1238 Mamalahoa Highway, Kamuela, HI 96743, USA
\and
Department of Astronomy, Columbia University, New York, NY 10027, USA
\and
Department of Physics and Astronomy, University of British Columbia, 6224 Agricultural Road, Vancouver, V6T 1Z1, BC, Canada
}

\date{accepted for publication in \aap}
\titlerunning{The VIPERS-MLS -- I: UV to NIR photometry and photometric redshifts}

\abstract{ We present observations collected in the CFHTLS-VIPERS region in the ultraviolet  with the GALEX satellite (far- and near-ultraviolet channels)  and in the near-infrared with the CFHT/WIRCam camera ($K_s$ band) over an area of 22 and 27 deg$^2$, respectively. The depth of the photometry was  optimised to measure the physical properties (e.g., star formation rate, stellar  masses) of all the galaxies in the VIPERS spectroscopic survey. The  large volume explored by VIPERS will enable a unique investigation of the relationship between the galaxy properties and their environment (density field and cosmic web) at high redshift ($0.5\le z\le 1.2$).  In  this paper, we present the observations, the data reductions, and the build-up of the multi-colour catalogues.  The CFHTLS-T0007  (gri-$\chi^2$) images are used as reference to detect and measure the $K_s$ -band photometry, while the T0007 $u^*$-selected sources are used as priors to perform the GALEX photometry based on a dedicated software (EMphot).  Our final sample reaches $NUV_{AB}\sim$ 25 (at 5$\sigma$) and  $K_{AB}\sim22$ (at 3$\sigma$).  The large  spectroscopic sample ($\sim$51,000 spectroscopic redshifts) allows us to highlight the robustness of our star/galaxy separation and the reliability of our photometric redshifts with a typical accuracy of $\sigma_z \le 0.04$ and a fraction of catastrophic failures $\eta \le$ 2\% down to $i \sim$ 23.  We present various tests on the $K_s$ -band completeness and photometric redshift accuracy by comparing our results with existing overlapping deep photometric catalogues. Finally, we discuss the BzK sample of passive and active galaxies at high redshift and the evolution of galaxy morphology in the $(NUV-r)$ vs $(r-K_s)$ diagram at low redshift ($z\le 0.25$) based on the high image quality of the CFHTLS.  The images, catalogues, and photometric redshifts for 1.5 million sources (down to $NUV \le 25$ $\cup$ $K_s\le 22$) are released and available at this URL: \url{http://cesam.lam.fr/vipers-mls/}.
 }
\maketitle

\section{Introduction}

The gain in sensitivity and resolution of modern instruments has given  access to the whole electromagnetic spectrum, allowing observations of the stellar light (thermal emission from the far-ultraviolet to the mid-infrared) and of the re-processed (dust in far-infrared / radio) and non-thermal (supernovae [SN], active galactic nuclei [AGN] in radio, mid-infrared and X-ray) emissions of distant galaxies.  In the past decade a wealth of multi-wavelength data has been collected in the so-called deep fields to measure galaxy physical properties such as stellar mass, star formation activity, dust content, and AGN fraction: GOODS \citep{Giavalisco2004}, VVDS \citep{LeFevre2005}, COSMOS \citep{Scoville2007}, DEEP2 \citep{Newman2013} or SXDS \citep{Furusawa2008} among others.
A clear picture has emerged for the global evolution of the star formation rate (SFR) and stellar mass densities over cosmic time \citep[e.g.,][]{Madau2014}.   We observe a general decline of the star formation rate density since $z\sim2$, a rapid build-up of the massive end of the mass function ($z\sim 1$), accompanied by the emergence of quiescent galaxies. However, the physical processes responsible for these evolutions are still to be understood.
 The gradual decline of star formation (SF) activity could be due to the dwindling gas  supply in galaxies, suppressing the fuel for star formation activity \citep[in line with the slowing growth rate of massive dark matter halos, e.g., ][]{Bouche2010}.  At high redshift, hydrodynamical simulations suggest that cold streams \citep[]{Katz2003} can penetrate deep into the dark matter halos and feed galaxies with fresh gas, allowing intense star formation activities and rapid formation of massive galaxies. Internal feedback processes  \citep[SN, AGN, virial shock heating;][]{Cattaneo2006, Keres2005} can then expel the gas and/or halt  the SF activity and produce the early formation of massive quiescent galaxies. Alternatively, galaxies may also become quiescent as a result of environmental effects such as galaxy harassment \citep[e.g.,][]{Farouki1981}, ram-pressure stripping \citep[][]{Gunn1972}, galaxy strangulation \citep[][]{Larson1980}, or major merging.  
  All these processes must be addressed in the global framework of large-scale structure formation and the complex filamentary network of the cosmic web, containing 90\% of the baryons, in order to evaluate the relative contribution of environmental effects (nurture) versus internal effects (nature) in shaping the galaxy properties observed today.  

 To analyse the relationship between galaxy properties and their environment, the latter has been defined in different ways: {\it i)} galaxy clustering analyses provide a statistical link between the galaxies and their host dark matter haloes \citep[][]{kauffmann2004}, which, combined with their abundances, can constrain the dark
matter halo occupation distribution \citep[]{zehavi2005};
 {\it ii)} the reconstruction of the local density field as measured with spectroscopic \citep[]{Cucciati2006, Cooper2008, Kovac2010} or photometric \citep[]{Scoville2013} redshifts allows high- and low-density contrast regions to be separated; {\it iii)}  the direct measurement of pair, group, and cluster membership can be compared to isolated field galaxies \citep[]{Haines2008, Wong2011, Peng2010, Kovac2014}; and {\it iv)} the reconstruction of the large-scale structure network composed of filaments, nodes, walls, and voids \citep[]{AragonCalvo2010b, Sousbie2011}. It offers a new way to describe the anisotropic environment of the cosmic web.  As mentioned before, the cosmic streams can feed galaxies in fresh gas, they also advect angular momentum into the disk of galaxies, which can lead to the alignment of the spine of spiral galaxies with the large-scale filaments, as predicted by tidal torques theory \citep[]{Codis2015} and recently observed in the SDSS survey \citep[]{Tempel2013}.

All these investigations are performed extensively in the local Universe with the help of the large spectroscopic redshift surveys, SDSS, 2dF, and GAMA.  At higher redshift, current surveys do not allow reconstructing the cosmic web because of either the small field of view or the low spectroscopic sampling rate. 
This situation has changed with the large VIMOS Public Extragalactic Redshift Survey \citep[VIPERS;][]{Guzzo2014} carried at ESO Very Large Telescope (VLT). With $\sim$100,000 spectroscopic redshifts over 24 deg$^2$ down to $i^*_{AB} < 22.5$ in
the redshift range $0.5 < z < 1.2$ \citep[]{Garilli2014}, VIPERS probes a volume equivalent to the local 2dF survey, providing a unique high-z sample to beat the cosmic variance ($\sim 10\%$) that affects current surveys (80$\%$ for GOODS down to 25$\%$ for COSMOS at $z\sim 0.8$). Its spectroscopic sampling rate ($\sim 40\%$) ensures a sufficient sampling to measure the redshift-space distortions \citep[]{delatorre2013} and to reveal the cosmic web on a large scale.

The VIPERS spectroscopic survey benefits from the optical coverage of the CFHT Legacy Survey Wide (ugriz)  in the W1 and W4 fields. To derive reliable stellar masses and SFR for VIPERS galaxies, we have undertaken a complete near-infrared (NIR) follow-up in the two fields  and partial follow-up in near-ultraviolet (only W1), which are described in this paper. 
In the past ten years, stellar population synthesis (SPS) models have been extensively used to convert luminosity into physical properties using a template-fitting procedure \citep[e.g., see recent reviews by ][]{Walcher2011, Conroy2013}.  The wavelength coverage  is critical to reduce the uncertainties on those parameters,
however. The NIR domain samples the rest-frame optical part of the spectral energy distribution (SED) of distant galaxies. Without NIR,  stellar mass measurements are uncertain ($\sigma_M \sim 0.4$ dex), with strong systematics at $z>0.8$ that are caused by the degeneracy between age, metallicity, and extinction (Ilbert, private communication).  The vast majority of VIPERS galaxies are below $z<1.0,$ and the CFHTLS imaging  does not fully probe the rest-frame ultraviolet (UV) light emitted by massive stars, which is crucial for an SFR estimate.  The VIPERS area has been partially observed with the GALEX survey \citep{Martin2005b} in the far-UV (135-175 nm) and near-UV (170-275 nm). However, measuring the GALEX photometry is a complex problem because of the large point-spread function of the telescope (FWHM$\sim$ 5"). The standard SExtractor photometry \citep[]{Bertin1996}  delivered by the GALEX pipeline is not optimal for faint sources because of blending problems. It requires a dedicated software \citep[]{Conseil2011}, as discussed in this paper. 

 With the physical properties of the VIPERS spectroscopic sources derived from the multi-wavelength dataset  described in this paper, 
 \citet[]{Davidzon2013} measured the high-mass tail of the galaxy stellar mass function (GSMF) with unprecedented accuracy for both quiescent and star-forming galaxies from  $z=$1.3 to $z=$0.5 using the first 53608 spectroscopic redshifts from VIPERS (over 7.5 deg$^2$).  \citet[]{Davidzon2016} extended their GSMF analysis as a function of environment by splitting galaxies into high- and low-density regions (using a local 3D galaxy density measurement) and detected an evolution of the GSMF in high-density regions. In contrast, the GSMF remains nearly constant  in low-density regions.

In addition to the VIPERS spectroscopic sample, the photometric multi-wavelength  survey ($F/NUV\sim 25$, $ugriz\sim 24.5$, $K_s\sim 22$)  combines depth and area ($\sim$ 24 deg$^2$) and provides accurate photometric redshifts and stellar masses for almost one million galaxies below $z\le 1.5$ as well as an efficient separation between active and passive galaxies based on a $NUVrK$ colour diagram \citep[]{Arnouts2013}. \citet{Moutard2016b} exploited this large volume to follow the evolution of the GSMF over the past 10 Gyr for   massive galaxies, $M\ge 10^{10}M_{\odot}$. It extended the precursor work by \citet[]{Matsuoka2010} performed in a 55 $deg^2$ region of the Stripe 82  overlapping the UKIDSS-LAS K-band  \citep[]{Lawrence2007}, where they  constrained the growth history of the most massive galaxies ($log(M/Mo)>11$ since $z\sim 0.9$. \citet[]{Coupon2015} used the superb image quality of the CFHT images  to perform a combined analysis of galaxy clustering, galaxy-galaxy lensing\footnote{The galaxy shape is obtained with the CFHTLenS catalogues.} , and the GSMF to produce a new constraint on the relationship between galaxies and their host dark matter halos. 

 This paper is organised as follows:  we present the  WIRCam-$K_s$  and GALEX-UV  observations,  the data reduction and the creation of the multi-colour catalogue based on the CFHTLS optical catalogue from the final TERAPIX release (T0007).  We perform various tests on our photometry and completeness by comparing them with deeper overlapping datasets and simulations. We use the large spectroscopic sample to quantify the quality of our photometric redshifts. As direct applications, we present the selection of a bright  BzK sample,  and we illustrate the morphology of low-z ($z\le 0.25$) large galaxies in the $(NUV-r)$ vs $(r-K)$ rest-frame colour diagram. 
Throughout the paper we adopt the following cosmology: $H_0$=70\ km
s$^{-1}$ Mpc$^{-1}$ and $\Omega_M=0.3$, $\Omega_{\Lambda}=0.7$. We
adopt the initial mass function of \citet{Chabrier2003}, truncated at
0.1 and 100 \Msun. All magnitudes are given in the AB system \citep{Oke1974}.
   
%

\begin{figure}
        \centering
\includegraphics[width=0.99\hsize, trim = 0cm 0.5cm 0cm 0cm, clip]{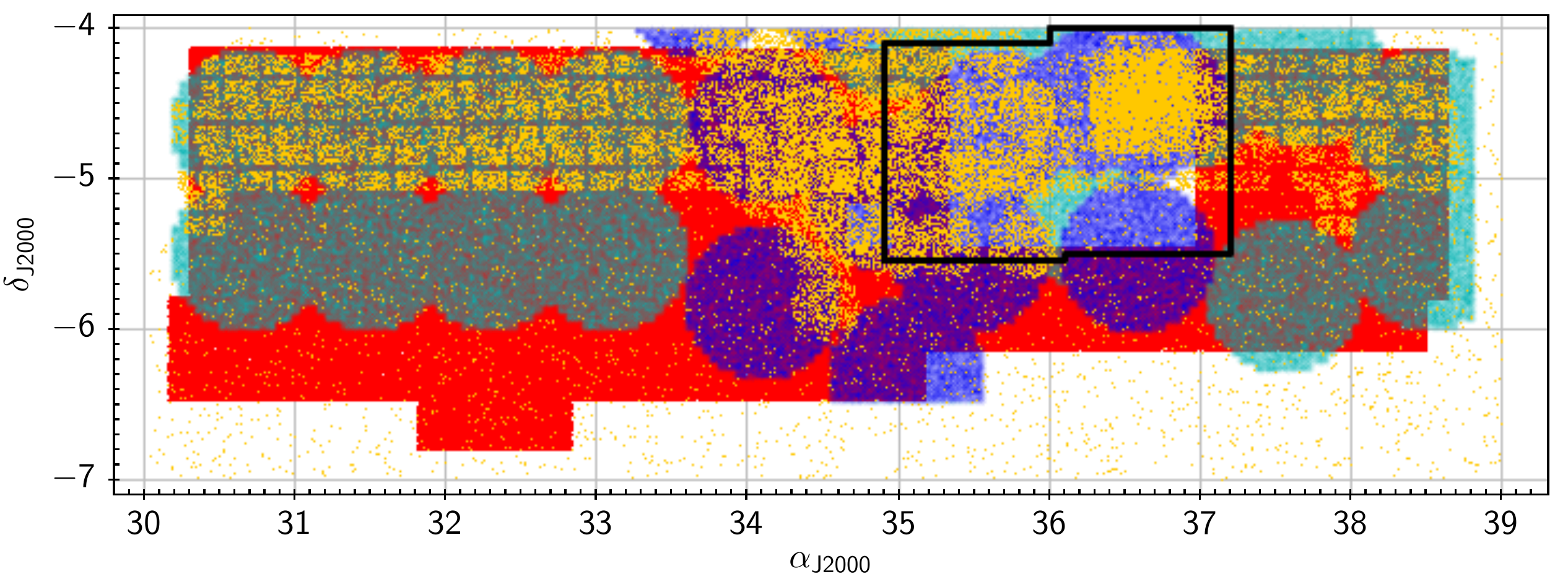}
\includegraphics[width=0.99\hsize, trim = 0cm 0cm 0.1cm 0cm, clip]{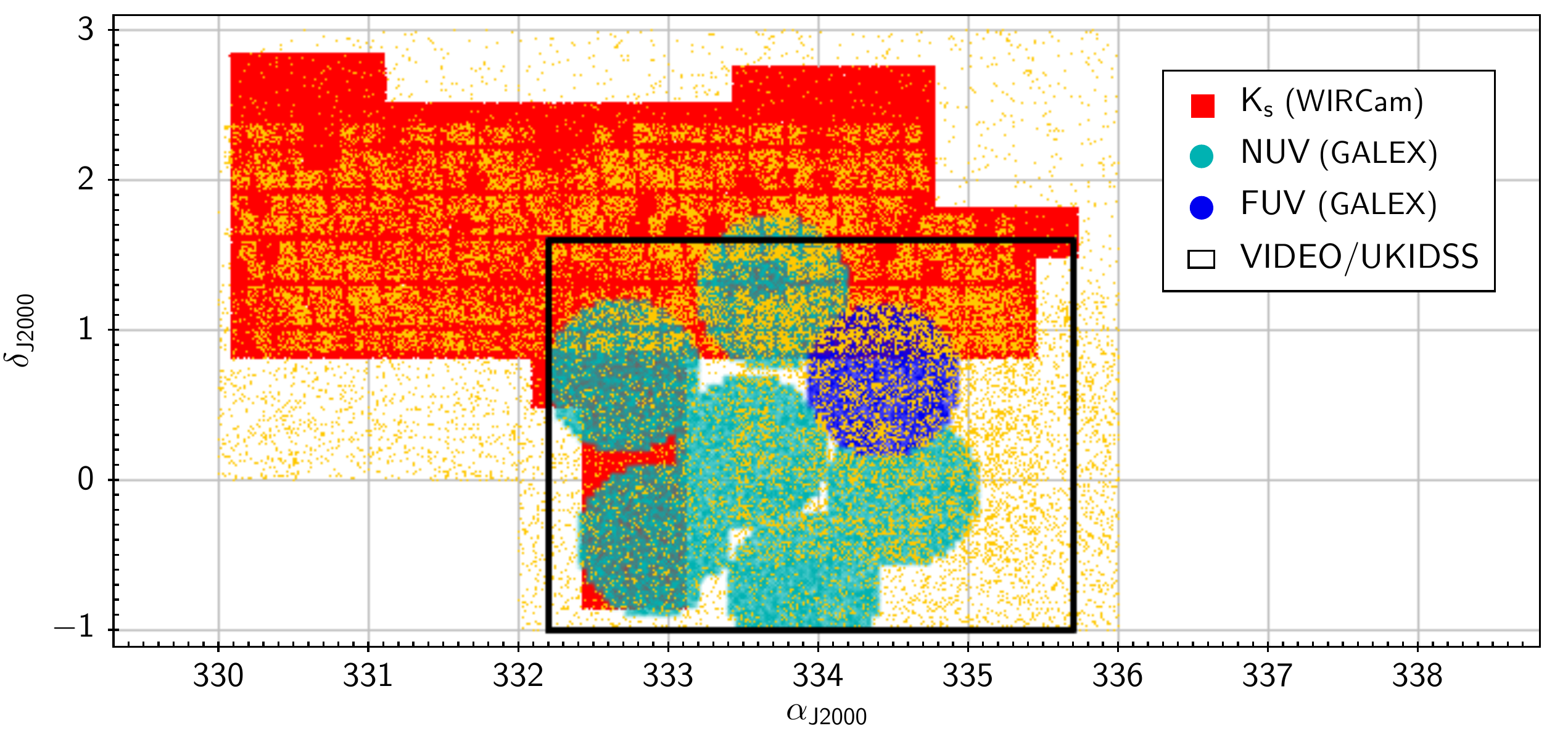}
\caption[]{Footprints of the GALEX NUV/FUV (cyan/blue) and WIRCam K band (red background) observations  in the CFHTLS W1 (top) and W4 (bottom) fields. The spectroscopic sources are shown as golden points. 
The total area (W1 and W4) is $\sim$7.8 and $\sim$22$\deg^2$ in FUV and NUV bands observed with GALEX and  $\sim$ 27 deg$^2$ with WIRCam. 
 The black layouts represent the VIDEO (W1) and UKIDSS (W4) surveys.
} 
\label{fig:layout}
\end{figure}
%

%
\section{Data}
\subsection{Optical CFHTLS photometry}
The Canada-France-Hawaii Telescope Legacy Survey\footnote{http://www.cfht.hawaii.edu/Science/CFHTLS/} (CFHTLS) is an imaging survey performed with MegaCam \citep{Boulade2000} in five optical bands $u^{\star}, g, r, i, z$.  The CFHTLS-Wide covers   
four independent patches in the sky over a total area of 154~deg$^2$ with sub-arcsecond seeing (median  $\sim 0.7"$) and a typical depth of  $i\ab \sim24.8$ (5$\sigma$ detection in 2$\arcsec$ apertures). In this study we use the images and photometric catalogues of the W1 and W4 fields from the worldwide T0007\footnote{http://terapix.iap.fr/cplt/T0007/doc/T0007-doc.html} release produced by TERAPIX\footnote{http://terapix.iap.fr/}. This final release includes an improved absolute and internal photometric calibration based on the  photometric calibration scheme adopted by the Supernova Legacy Survey (SNLS) and described  in \citet[][]{Regnault2009}.  The main ingredients are  {\it i)} the use of the  spectrophotometric standard star BD +17 4708 instead of the Landolt standard stars; {\it ii)}  a new photometric flat-fields used in the reduction pipeline by Elixir at CFHT on the raw images, which guarantees a flat photometry across the MegaCam field of view; and {\it iii)} the use of shallow photometric calibration observations covering the whole CFHTLS wide in each filter and bracketed by SNLS tertiary standards. These steps bring the absolute and relative photometric calibrations over the entire survey to an accuracy level of between 1 to 2\%.
 
The final images were stacked with the software SWARP
\citep[][]{Bertin2006}. The photometry was performed with
SExtractor \citep[][]{Bertin1996} in dual mode with the source
detection based on the $gri-\chi^2$ image \citep{Szalay1999}.
Following \citet[]{Erben2013}, we used the isophotal apertures,
extracted from the detection images, to measure the galaxy colours. The apertures are smaller than the Kron-like apertures
\citep[]{Kron1980}, which provides less noisy estimates of the colours and
leads to an improved photometric redshift accuracy
\citep{Hildebrandt2012}.
 
To derive the total magnitudes, we rescaled the isophotal magnitudes
(\texttt{MAG\_ISO}) to the quasi-total magnitudes (\texttt{MAG\_AUTO})
for each source.  The scaling factor, $\delta_m$, combines the
individual scalings in g', r', i', and $K_s$ band, when available (the
$K_s$ -band images share the same pixel grid as the optical images, see
Sect.~\ref{sec:nir}), to account for the intrinsic colours
of galaxies and their respective uncertainties. It is defined as
\begin{equation}
\label{eq:dm}
\delta_m =  \dfrac{  \sum_f  (m_{\it AUTO}- m_{\it ISO})^f  \times w^{f}}{ \sum_f  w^f} 
,\end{equation}
where f corresponds to the  g', r', i', and $K_s$ filters, and the weight, $w^f$, is defined as $1/w^f = (\sigma_{\it AUTO}^2+\sigma_{ISO}^2)^f $.
 The final magnitudes for each source are then simply defined by a unique shift applied to all passbands:
$m^f =  m_{\it ISO}^f + \delta_m$.

  Finally, we also investigated the best masking procedures of the bad regions by comparing the masks produced by TERAPIX and the CFHTLenS team\footnote{http://www.cfhtlens.org/}. After visual inspections, we found that CFHTLenS masks provide a better cleaning of spurious objects around bright stars by adopting a large central disk region on top of the elongated spikes \citep{Erben2013}. We therefore decided to adopt their more conservative masking approach. 
   
  In conclusion, our CFHTLS optical catalogue is similar to the original T0007 in terms of source detections, but differs in the masking regions and the choice of magnitudes, where we adopted a scheme that allows optimizing the colour estimates for the photometric redshifts  and the total flux for the measurements of the physical parameters. In Fig. \ref{fig:layout} UV (GALEX) and NIR (WIRCam $K_s$ -band) observations overlap the CFHTLS. 

\subsection{NIR observations}
\label{sec:nir}
\subsubsection{Observations}
\begin{figure}
\includegraphics[width=9cm,height=8cm]{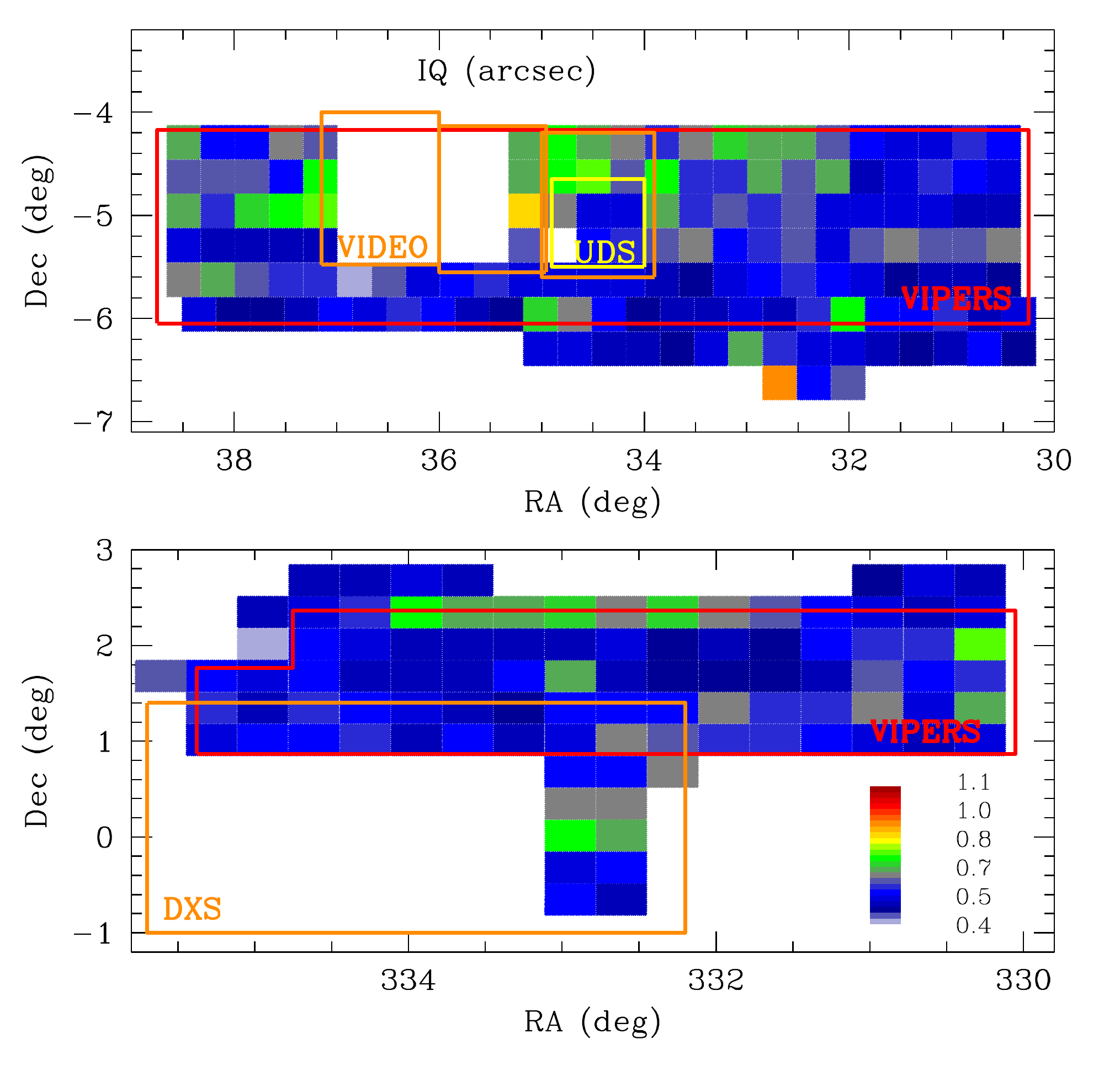}   
\caption[]{ Colour-coded mean image quality (IQ) per WIRCam tile over the whole WIRCam survey (W1: top and W4: bottom). The layouts indicate the VIPERS regions (red layouts), the three VIDEO pointings  (orange in W1), the UKIDSS-UDS (yellow in W1), and UKIDSS-DXS (orange in W4).}
\label{fig:iq}
\end{figure}

Since 2010, we have conducted a $\Ks$-band follow-up of the VIPERS
fields with the WIRCam instrument \citep{Puget2004} at CFHT. The
original motivation was to guarantee an almost complete detection in
$K_s$ band of the VIPERS spectroscopic galaxies ($i_{AB}\le 22.5$). To
do so, we estimate that we need to detect sources down to $K_s\sim22$
(at $\sim$ 2-3$\sigma$), which can be reached with a minimum
integration time per pixel of $\sim$1050 seconds.  The observations
have been executed with two separated observing groups with 21
exposures of 25sec each. The WIRCam observations are shown in red in Fig.\ref{fig:layout}. They cover a total area of 27deg$^2$ and correspond to a total amount of allocated
time of $\sim$120 hours spread during the period 2010 and 2012.  As a result of high sky background or bad seeing ($IQ>0.8\arcsec$), a few
observing groups have been repeated. The average seeing for each
WIRCam tile is shown in Fig.~\ref{fig:iq} for the two fields.  The
mean image quality is very homogeneous and corresponds to $<IQ> =
0.6\arcsec \pm$0.09. We note that the missing region in the W1 field
corresponds to the location of the VIDEO survey. The southern
extension of W4 outside the VIPERS regions was planned to cover the
GALEX field that was initially not covered by the earliest release of the
UKIDSS DXS survey.

\subsubsection{Data reduction}
\begin{figure}
\centering
\includegraphics[width=8.5cm,height=8cm]{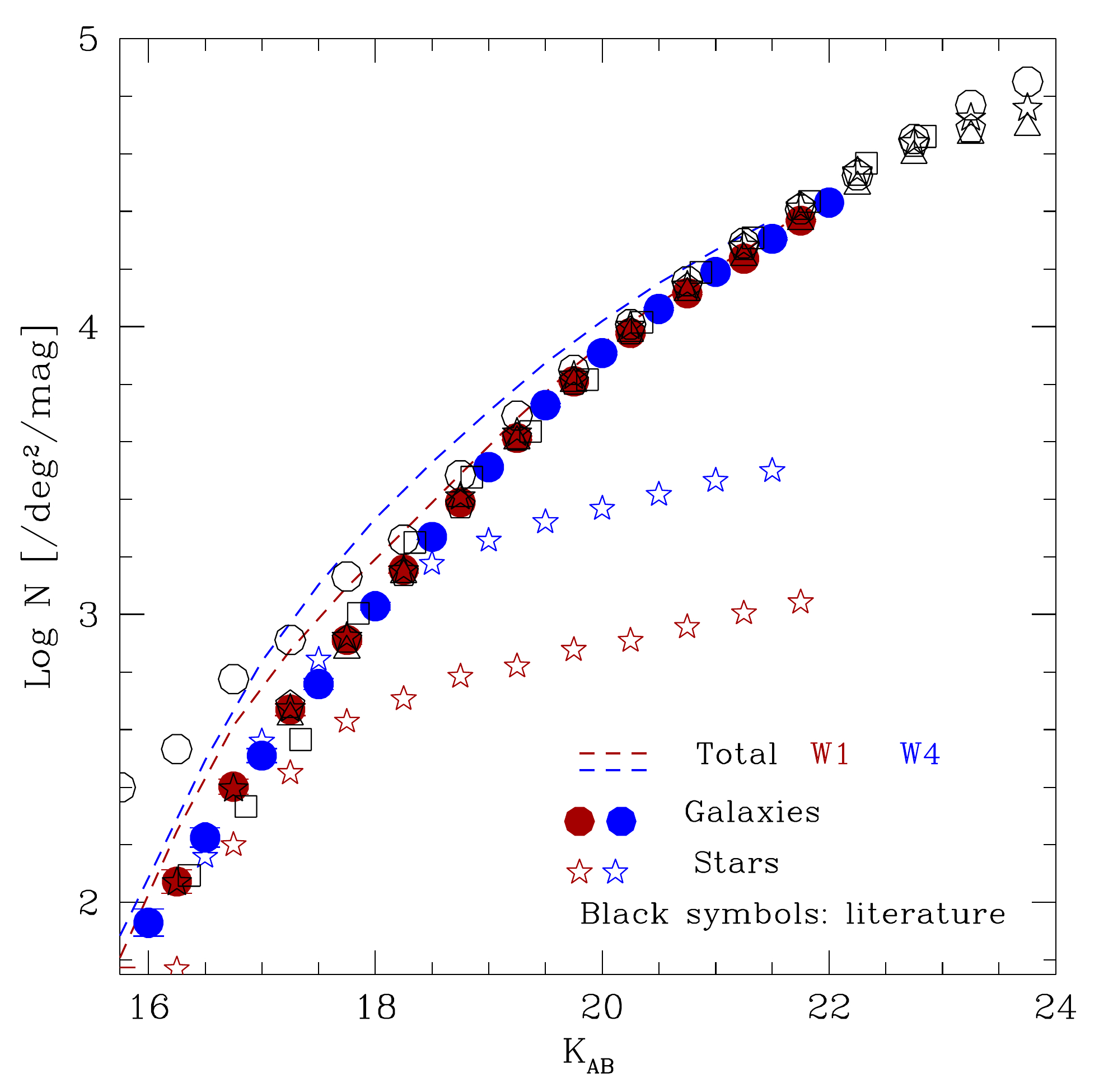}   
\caption[]{ WIRCam $K_s$ number counts for the galaxies (filled circles) and stars (open stars) in the W1 (red) and W4 (blue) regions. The dashed lines show the total number counts.  The open dark symbols are a compilation from the literature (triangles: \citet[]{McCracken2012};  pentagons: \citet[]{Jarvis2013};  circles: UDS (DR8);  squares:\citet[]{quadri2007};  stars: \citet[]{Bielby2012}    }
\label{fig:Nk}
\end{figure}

All the data were reduced at CFHT and TERAPIX.
The initial processing with the i'wii pipeline\footnote{http://cfht.hawaii.edu/Instruments/Imaging/WIRCam/\\ IiwiVersion1Doc.html} at CFHT removes the instrumental imprints on the individual images (flagging of saturated, bad, and hot pixels, correcting for nonlinearity, bias removal, dome and sky flat-fielding, and guide window masking). 
 CFHT delivers detrended images with a first astrometric solution, photometric zeropoint, and a sky subtraction for each exposure. \\
   TERAPIX first uses QualityFITS on detrended images to produce weight-maps (with the WeightWatcher software; \citet{Marmo2008}), source catalogues and checks the individual image qualities (e.g., seeing and depth). The validated exposures are then median-combined to produce an initial stack by using the  Scamp and Swarp softwares\footnote{ http://www.astromatic.net} \citep{Bertin2006}.  To improve the sky subtraction and the photometry of faint sources, a two-step strategy is adopted similar to the CFHT-WIRDS \citep{Bielby2012} and the ESO-UltraVISTA \citep{McCracken2012} surveys. 
 The initial stack is used to produce a mask with all the faint sources.
 In a second pass, the mask is applied to the individual exposures and a new sky background is estimated  by medianing images inside a sliding window and for images within a time window $\Delta$t $<$ 15 min and angular separation $\Delta \theta < 10 \arcmin$. 
 A last quality assessment of the individual sky-subtracted images is performed before producing the final stacks.
  The final WIRCam products  are composed of  stack images and weight images resampled on the footprint of the CFHTLS-MegaCam tiles and with the same pixel scale ($0.18\arcsec$/pix) using
the SWARP software with a Lanczos3 resampling kernel. Like for the T0007 release, the photometry is performed in dual-image mode with the $gri-\chi^2$ image and the same \texttt{SExtractor} setting \citep{Hudelot2012}. 
  A last correction is applied to the SExtractor error measurements, which are underestimated as a result of the noise correlation introduced by the stacking or resampling process. To estimate this factor, we also performed stacks on the native WIRCam grid (and original pixel-scale) with a nearest-neighbour technique to preserve the white-noise nature in the final images. When we
compared the $K_s$ uncertainties of the same  sources, we found
that a factor 1.5 needed to be applied to the error estimates with the megaCam grid.   No masking other than the CFHTLenS masks needed to be applied to the WIRCam sample. The effective WIRCam $K_s$ area, after applying the masks, drops from 27 deg$^2$ to  22.4 deg$^2$ (including W1 and W4 fields). The WIRCam number counts in the two fields are shown in Fig.~\ref{fig:Nk} separately for stars and galaxies (see Sect.~\ref{sec:sg}) and are compared with literature. 
%

%
%
%
\begin{figure}
\includegraphics[width=8.5cm,height=8cm]{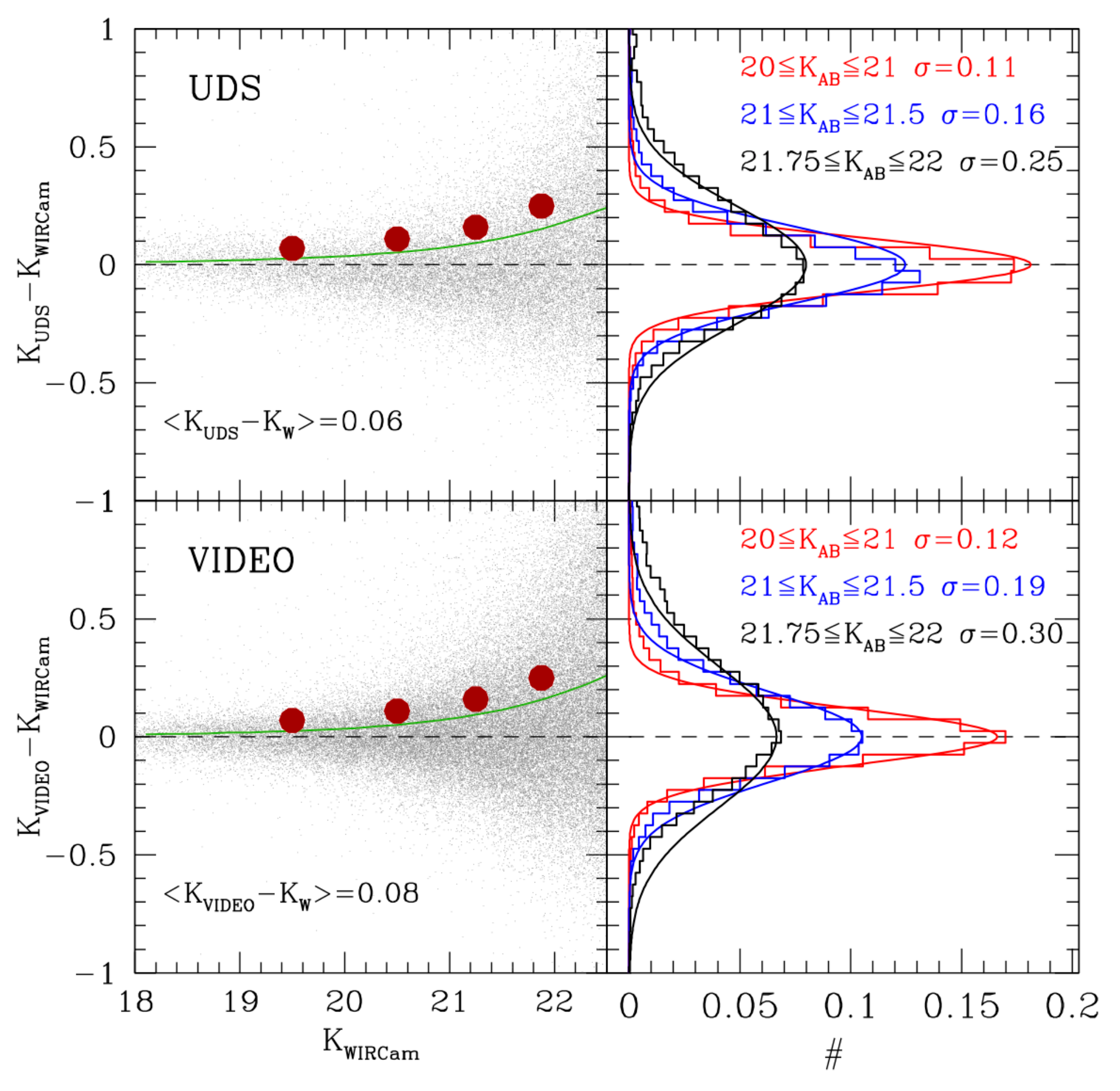}   
\caption[]{Comparison of the VIPERS K-band photometry with the
  Ultra-Deep UKIDSS field (top left), the VIDEO XMM2 field (bottom
  left), and the scatter distributions in three magnitude bins (right
  panels). The observed scatters are reported in the
  left panels as red circles, while the green solid lines show the mean
  magnitude vs error behaviour of our VIPERS K band data after
  rescaling the original errors by 1.5 (see text).  
  Larger dispersions are observed with VIDEO than with UDS 
  because of the different depths between the two fields. 
}
\label{fig:dk}
\end{figure}
%
\subsubsection{Comparison with deeper datasets} 
As shown in Fig.~\ref{fig:iq}, we can use the existing deep near-IR surveys, UKIDSS-Ultra Deep Survey \citep[UDS]{Lawrence2007} and VIDEO \citep{Jarvis2013}, to estimate the photometric accuracy, depth, and completeness of our $K_s$ sample.  UDS and VIDEO are significantly deeper than our dataset ($K\sim 24.5$ and 23.8 respectively). For both fields the relative astrometric accuracy with our sample is $\sigma \le 0.09\arcsec$ with systematic shifts of the same order. We adopted a maximal matching distance of $0.5\arcsec$.  To compare the fluxes, we used  the SExtractor {\it MAG\_AUTO} magnitudes for the VIDEO survey and the petrosian magnitudes for the UDS survey (which should be close to our own magnitudes).  Figure \ref{fig:dk} shows the magnitude differences as a function of magnitude (on the left)  and the dispersions in three magnitude intervals (on the right)  after correcting for the systematic shifts ($K_{DEEP}-K_{WIRCam} \sim$0.06 and 0.10 mag for UDS and VIDEO, respectively).  For both surveys the mean difference is stable and close to zero down to $K_s \sim 22$, confirming that no bias due to sky-background subtraction is present in our data.  At all magnitudes, the dispersions are reasonably well fitted by a  Gaussian distribution.  We expect to reach  $\sigma \sim$0.3 mag (S/N$\sim 3$) near $K_s\sim 22$.  The dispersions predicted with these external dataset are consistent with those estimated by SExtractor (after the 1.5 rescaling factor mentioned above). 
%

\begin{figure}
\centering
\hspace{-0.2cm}\includegraphics[width=0.9\hsize]{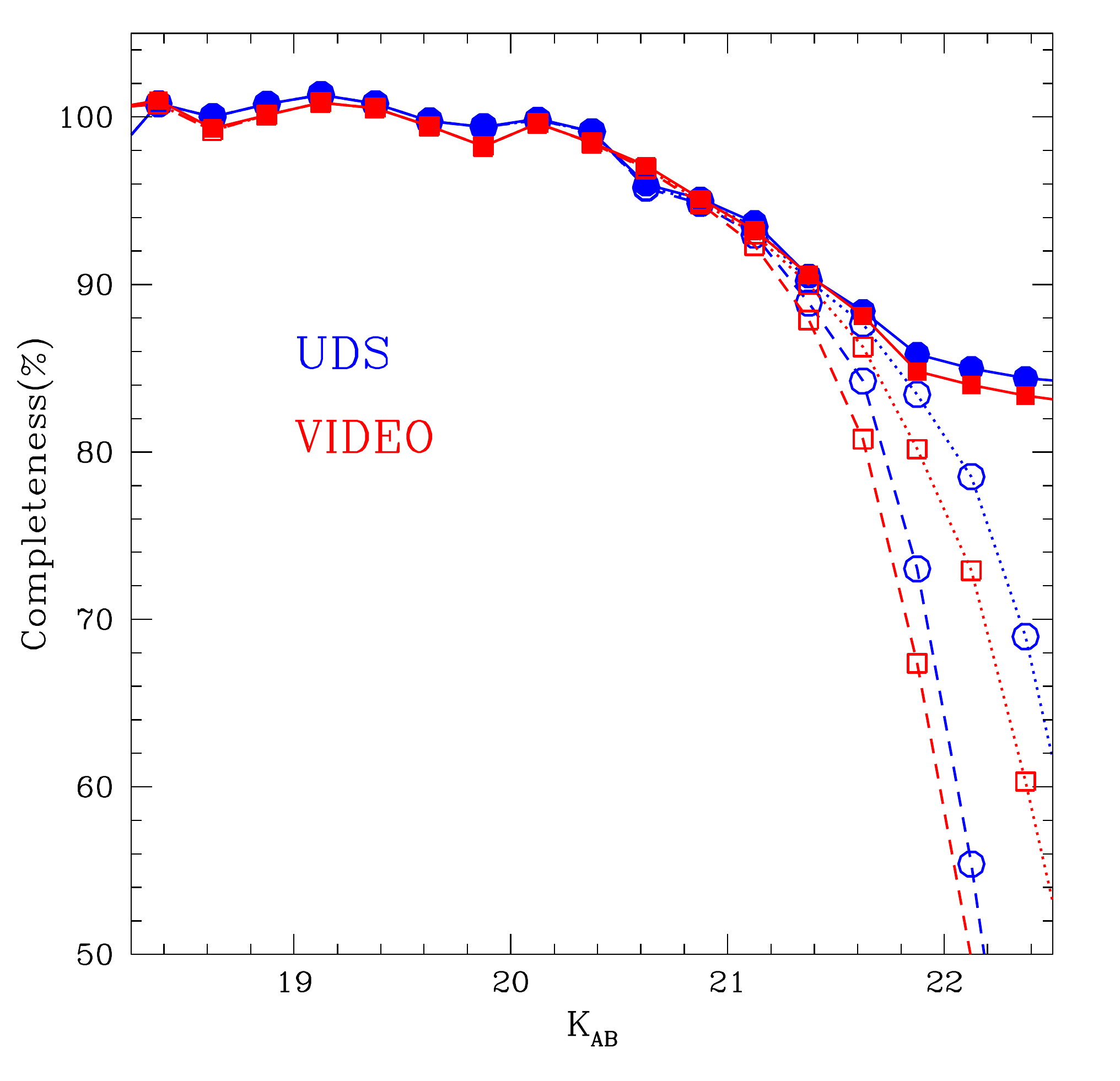}   
\includegraphics[width=0.9\hsize]{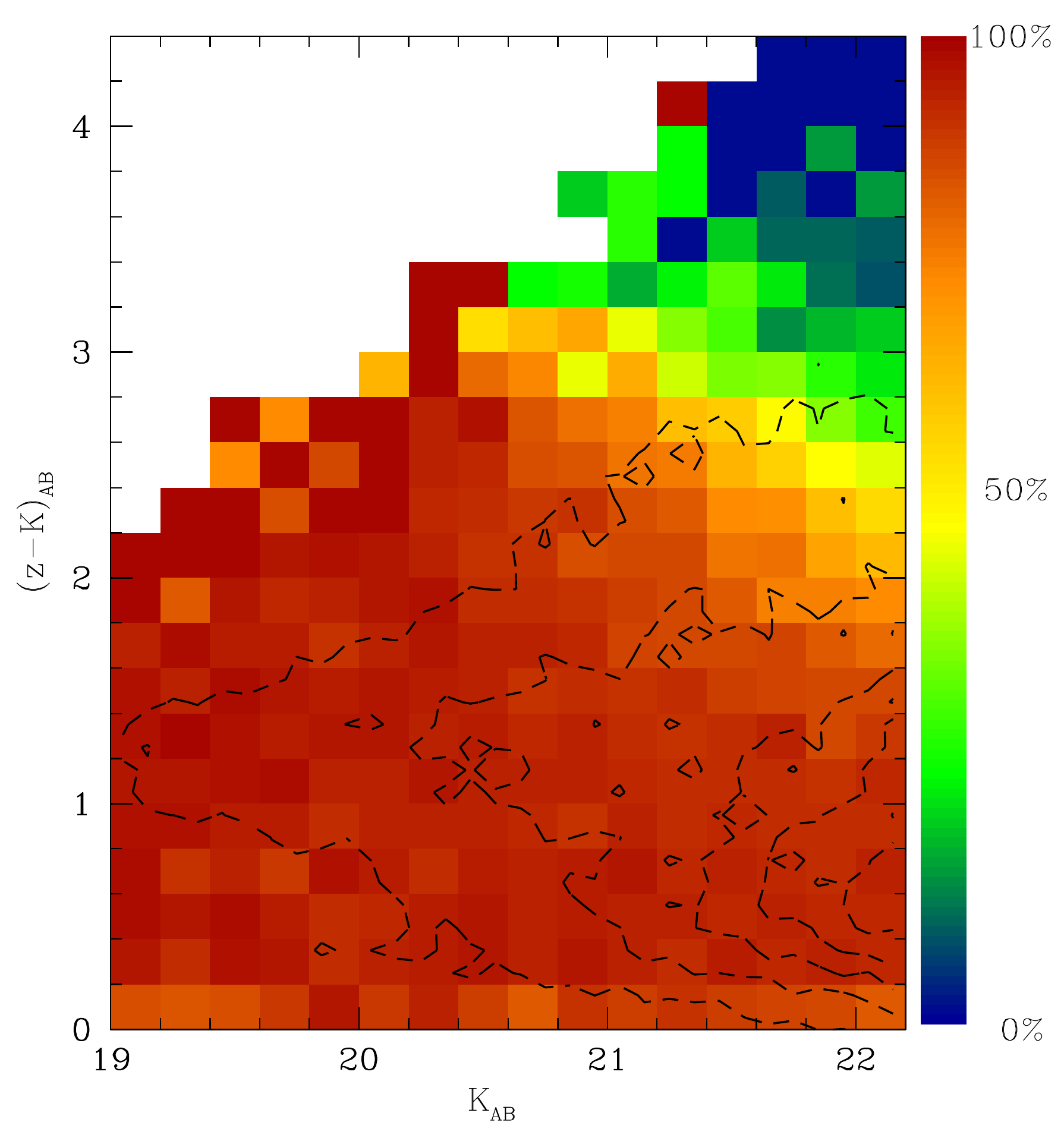}   
\caption[]{ Top panel: completeness (in \%) of the WIRCam sources as a
  function of magnitude, based on the deep surveys UDS
  (blue) and VIDEO (red). The completeness is shown for WIRCam
  sources with various signal-to-noise ratio cuts (no cut:  filled symbols;
   $S/N>$3: dotted lines ; $S/N>$5: dashed lines).   Bottom panel: completeness as a function of magnitude and colour $(z-K)$, colour-coded from red (100\%) to blue (0\%) based on the VIDEO survey.  The yellow region delineates the 50\% completeness level.  The dashed line contours show the source density spaced with logarithmic scale of 0.3 dex. At $K_{AB}\le$20.6, the WIRCam  sample is close to 100\% completeness and the colour variations are due to small numbers.  }
\label{fig:kcomp}
\end{figure}

%
Because we used the $gri-\chi^2$ images for the detection, our completeness is not directly related to the background noise of the WIRCam images.  To estimate the completeness, we used all the sources detected in the UDS and VIDEO images and measured the fraction of them detected in our optical-based catalogue.  In Fig.~\ref{fig:kcomp} (top panel) we show the completeness for the UDS (circles) and the VIDEO (squares) surveys as a function of magnitude. The solid lines and filled points refer to all the WIRCam sources irrespective of their signal-to-noise ratio
(S/N). Identical results are obtained in both fields, which shows that our catalogue detects more than 85\% of the  pure $K_s$ -band
selected surveys down to $K_s\sim 22$.   If we restrict the detection to WIRCam sources with  S/N$>$3 and 5, the completeness drops slightly to 81\% and 66\% (based on the UDS survey), respectively,  at $K_{AB}\sim 22$. This confirms that our gri-based   catalogue
is almost complete down to the $K_{AB}\sim22$ magnitude.

We can investigate the nature of this incompleteness in more detail, which should preferentially reflect a bias against the reddest sources.
To do so, we measured the completeness as a function of $K$ magnitude and
$(z-K)$ colour for the VIDEO survey alone (which includes its own z-band photometry that reaches a depth $z\sim 25.7$ at 5$\sigma$). 
The bidimensional completeness is shown in Fig.~\ref{fig:kcomp} (bottom panel).
 The colour-coding reflects the completeness level, with yellow corresponding to a 50\% completeness. Below $K_{AB}\sim 20.6$, the catalogue is complete regardless of $(z-K)$ colour. 
 At fainter $K$ sources, the completeness of the reddest sources
decreases gradually. For this reason, any statistical analysis such as luminosity functions or stellar mass functions, based on a $K$ selection, should apply a weight reflecting this bidimensional completeness map.

\subsection{UV photometry}

\subsubsection{Observations}

Launched in 2005, the GALEX satellite \citep{Martin2005b} has explored
the ultraviolet sky in FUV (135-175 nm) and NUV (170-275 nm) passbands
\citep{Morrissey2005} and performed all-sky medium- and deep-imaging surveys.  The UV continuum is a direct indicator (except for the
dust) of the instantaneous SFR.  To
assess the continuum of high redshift galaxies, we here only considered the
observations from the Deep Imaging Survey (DIS, with $T_{\rm exp} \ge
30$~ksec).  All the GALEX pointings are shown in Fig.~\ref{fig:layout}
as blue (FUV) and cyan (NUV) circles.
  To complete the coverage of the VIPERS-W1 field in the NUV channel (after the shut-off of the FUV channel), we observed new pointings in the western and eastern parts of the field with $\sim$100 hr of discretionary time. Only the GALEX PanStarr fields, centred on the VVDS-22hr region, are available in the W4 field, with a small overlap with the VIPERS-W4 area.   All the GALEX pointings and their respective integration times are given in Table~\ref{tab:galex}. The total area observed with GALEX (W1 and W4) is $\sim$7.8 and $\sim$22deg$^2$ in the FUV and NUV bands.

\begin{table}[htdp]
\begin{center}
\caption{List of GALEX fields with integration times and central positions}
\begin{tabular}{|l|l|l|l|l|}
\hline 
 GALEX field & \multicolumn{2}{c|}{T$_{exp}$(ks)} & RA &  Dec \\  
                      & NUV & FUV& (deg) & (deg)  \\ 
\hline
 \multicolumn{5}{|c|}{W1 field}\\
         CFHTLS\_W1\_MOS00   & 27 & ---  & 38.5   & -4.65  \\
         CFHTLS\_W1\_MOS01   & 29 & ---  & 38.5   & -5.5          \\
         CFHTLS\_W1\_MOS02   & 30 & ---  & 37.7   &   -4.28 \\
         CFHTLS\_W1\_MOS03   & 29 & ---  & 37.55 & -5.78          \\
         CFHTLS\_W1\_MOS04   & 28 & ---  & 33.1   & -4.65  \\
         CFHTLS\_W1\_MOS05   & 29 & ---  & 33.1   & -5.5          \\
         CFHTLS\_W1\_MOS06   & 29 & ---  & 32.3   & -4.65  \\
         CFHTLS\_W1\_MOS07   & 26 & ---  & 32.3   & -5.5          \\
         CFHTLS\_W1\_MOS08   & 30 & ---  & 31.5   & -4.65  \\
         CFHTLS\_W1\_MOS09   & 30 & ---  & 31.5   & -5.5          \\
         CFHTLS\_W1\_MOS10   & 32 & ---  & 30.7   & -4.65  \\
         CFHTLS\_W1\_MOS11   & 33 & ---  & 30.7    & -5.5         \\
         PS\_CFHTLS\_MOS02    & 23 & --- & 34.95  & -3.96  \\
         PS\_CFHTLS\_MOS03    & 32 & --- & 35.87  & -4.25  \\
         PS\_CFHTLS\_MOS04    & 35 & --- & 36.9    & -4.42  \\
         PS\_CFHTLS\_MOS05    & 33 & --- & 35.2    & -5.05  \\
         PS\_CFHTLS\_MOS06    & 34 & --- & 36.23  & -5.2          \\
         XMMLSS\_00                     & 75 & 60 & 36.66  & -4.48  \\
         XMMLSS\_01                     &115& 92 & 36.36  & -4.48  \\
         XMMLSS\_02                     & 27 & 27 & 36.58  & -5.51  \\
         XMMLSS\_03                     & 30 & 29 & 35.47  &    -5.51  \\
         XMMLSS\_04                     & 29 & 27 & 34.91  & -5.1      \\
         XMMLSS\_05                     & 26 & 26 & 36.48  & -3.60 \\
         XMMLSS\_06                     & 29 & 29 & 34.6    & -3.65 \\
         XMMLSS\_07                     & 28 & 27 & 34.1    & -5.82 \\
         XMMLSS\_08                     & 31 & 31 & 35.05  & -6.29  \\
         XMMLSS\_09                     & 32 & 31 & 34.11  & -4.64 \\
         XMMLSS\_12                     & 22 & 21 & 33.66  & -3.68 \\
         XMMLSS\_20                     & 27 & 26 & 35.65  & -4.65 \\
\hline
 \multicolumn{5}{|c|}{W4 field}\\
 VVDS22H                          & 79        & 76       & 334.4  & 0.67 \\
 PS\_VVDS22H\_MOS00   & 54      &  --- & 333.7  & 1.25 \\
 PS\_VVDS22H\_MOS01   & 53      &  --- & 332.7  &  0.7 \\
 PS\_VVDS22H\_MOS02   & 59      &  --- & 334.4 &  0.67 \\
 PS\_VVDS22H\_MOS03   & 38      &  ---  & 333.5  & 0.18 \\
 PS\_VVDS22H\_MOS04   & 37      &  ---  & 334.5  & -0.05 \\
 PS\_VVDS22H\_MOS05   & 52      &  ---  & 333.9  & -0.72 \\
 PS\_VVDS22H\_MOS06   & 52      &  ---  & 332.9  & -0.4   \\
 \hline
\end{tabular}
\end{center}
 \label{tab:galex}
\end{table}%

\subsubsection{UV photometric extraction}

The data were reduced with the standard GALEX pipeline. In
the specific case of the DIS observations, the large PSF
(FWHM$\sim 5\arcsec$) affects the extraction and photometry of faint UV
sources because sources are confused \citep{Xu2005}.  As an alternative
to the source extraction of the GALEX pipeline (based on SExtractor
software), we developed a dedicated code to perform the UV
photometry, \texttt{EMphot} \citep{Guillaume2005, Conseil2011}.
EMphot uses a Bayesian approach with optical priors based on the
CFHTLS (T0007) $u^*$ -selected sources.  A full description of the code
and extensive simulations will be given in a forthcoming paper;
we outline the main steps below.
 
We generate a simulated GALEX image combining all the optical priors
with a relative flux contribution that must be estimated. The priors can either be
considered as a simple Dirac function or as an extended 2D profile
extracted from the $u^*$ -band image (i.e., stamps), which is what we adopted here.  The formalism is the following:  each observed value of the
GALEX pixel ($x_i$) is considered as a sample of a random variable
$X_i$ with the expected value being $\mu_i = E\{X_i\}$. 
The value $\mu_i$ at pixel $x_i$ is the contribution of all the input optical priors  around it and the sky background. It can be defined as 
\begin{eqnarray}
\mu_i = \sum_{k=1}^K \alpha_k h_{k,i} + b_i  ~,
\end{eqnarray}
 where $b_i$ is the background level value,  $h_{k,i} =\sum_j o_{k,j}f_{i-j}$ is the sum of all the objects, $o_{k}$, convolved by the GALEX PSF, $f$. The function $h_{k,i}$  is normalised to unity and  a set of scaling factor, $\alpha_k$, is applied to each prior $k$. 

 Here, we considered that the variable $X_i$ follows a Poisson statistic:  $P\{X_i=x_i\} = \exp(-\mu_i)\frac{\mu^{x_i}_i}{x_i!}$ . 

To estimate the $\alpha_k$ values, \cite{Guillaume2005} introduced the expectation maximisation scheme \citep[EM, see][]{Horiuchi2000}, which leads to  the iterative process
\begin{eqnarray}
  \alpha_k^{(n+1)} & = & \alpha_k^{(n)}
  \frac{  \sum_{i=1}^M (x_i/\mu_i^{(n)}) h_{k,i}}{\sum_{i=1}^M h_{k,i} }   \quad 
  ,\end{eqnarray}
  where $\mu_i^{(n)} = \sum_{j=1}^K \alpha_j^{(n)} h_{j,i} + b_i$. The E step compares the data image $ x_i $ to the projection $\mu_i^{(n)}$ of
the $\alpha_i^{(n)}$ estimates. The result is introduced in the M step as the
corrective ratio needed for the new set of $\alpha_k^{(n+1)}$ estimates. 
 Although the global system may take very many iterations to converge, we found that 100 iterations is sufficient to obtain stable $\alpha_k$ values for the main contributors, while priors with negligible values may oscillate.  To estimate the error of object k, 
we empirically quantify the variance $\sigma_k$, by the rms  under the object's area of the residual (best-fit model subtracted from the observed image) as  \\
\begin{equation}
\sigma_k^2 =  \frac{\sum_{i=1}^{M} h_{k,i}(x_i-\hat\mu_i)^2}{\sum_{i=1}^{M}h_{k,i}^2} ~.
\end{equation}
We find that this estimate is consistent with the errors obtained from simulations.

\begin{figure}
\includegraphics[width=8.5cm,height=6cm]{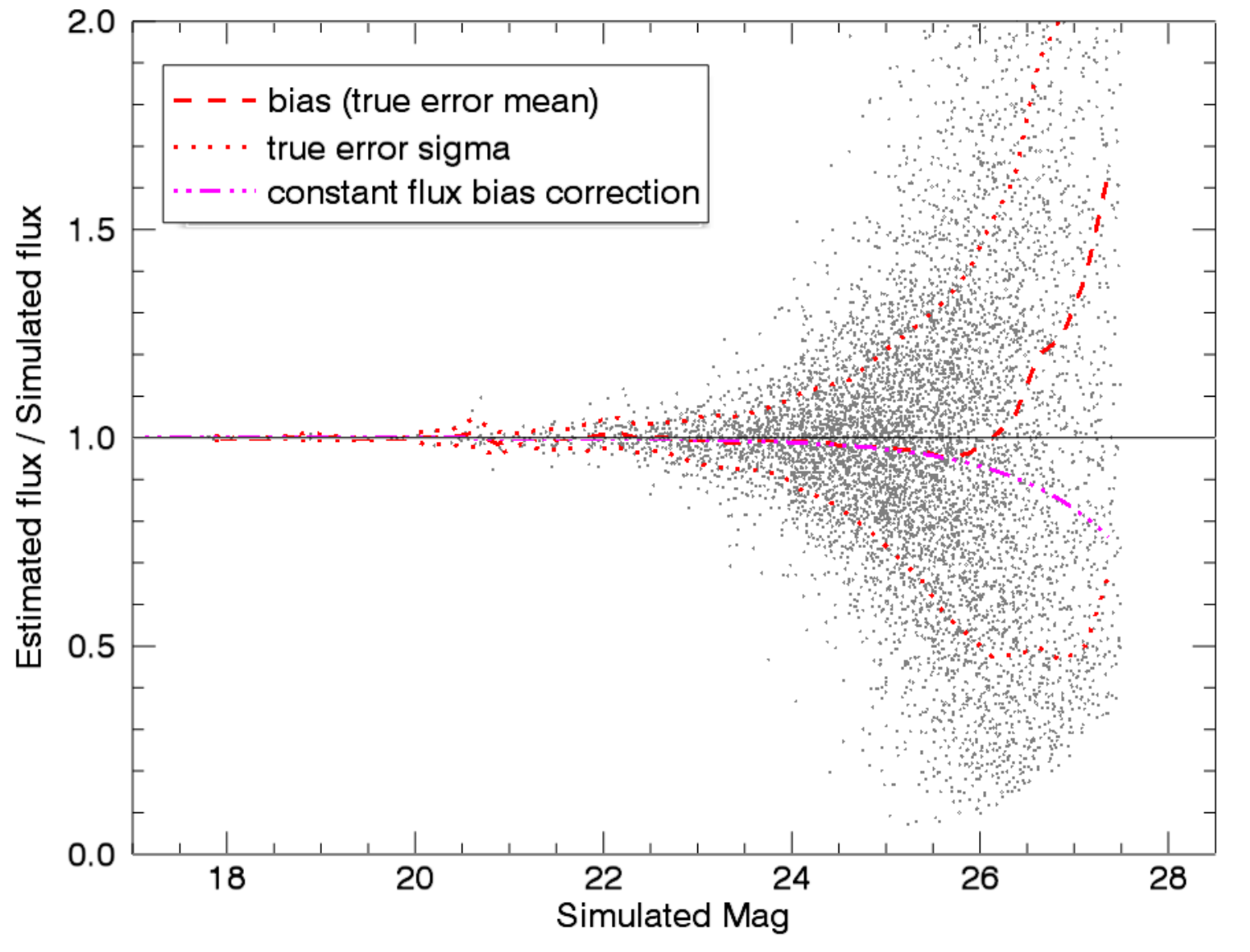}
\caption[]{ Comparison of the EM vs simulated fluxes as a function of simulated magnitudes. The dotted lines show the 1$\sigma$ error, the red dashed line the mean bias. The magenta dot-dashed line reflects the effect of a small background bias on the flux ratio. }
\label{fig:galex1}
\end{figure}

 We tested the EM photometric technique by adding multiple sets of simulated sources inside the observed GALEX images. In Fig.~\ref{fig:galex1} we show the flux ratio between the EM fluxes and the simulated fluxes as a function of magnitude.  The EM algorithm allows us to derive accurate UV fluxes down to NUV$\sim$25 with a S/N$\ge$5. This is consistent with the theoretical estimate of the S/N inside the PSF aperture and the measured background level. We observe a negligible mean bias (red dashed line) down to $NUV\sim 26,$ and its behaviour can be reproduced by a small residual background offset (magenta dot-dashed line).
%

\begin{figure}
\includegraphics[width=8.5cm,height=6cm]{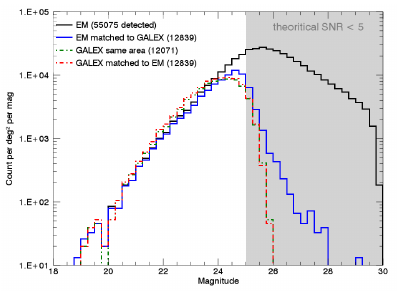}   
\caption[]{ NUV number counts vs magnitude in one GALEX field. The full EM photometric catalogue is shown as a solid black line and the EM sources matched to the GALEX pipeline sources as a
solid blue line. The GALEX pipeline counts  in the same area (and matched to EM sources) are shown as red and green dot-dashed lines.}
\label{fig:galex2}
\end{figure}

In Fig. \ref{fig:galex2} we show the NUV number counts from the EM software  (black line) and the GALEX pipeline  (green line) in one GALEX field. 
 First, the GALEX pipeline shows an excess of bright sources  ($22\le NUV\le 25$). This is explained by the deblending of the neighbouring objects in the EM technique,  which leads to systematically lower fluxes for EM. This is illustrated by comparing the number counts for the same objects as a function of GALEX  (red line) and  EM (blue line) magnitudes. 
 Second, the GALEX number count drops shortly after $NUV\sim 24$ (red line), while EM counts continue to increase down to $NUV\sim 25$. This is indeed expected because we used optical priors and because the sources were deblended.  
 
\begin{figure}
\centering
\includegraphics[width=8.5cm,height=8cm]{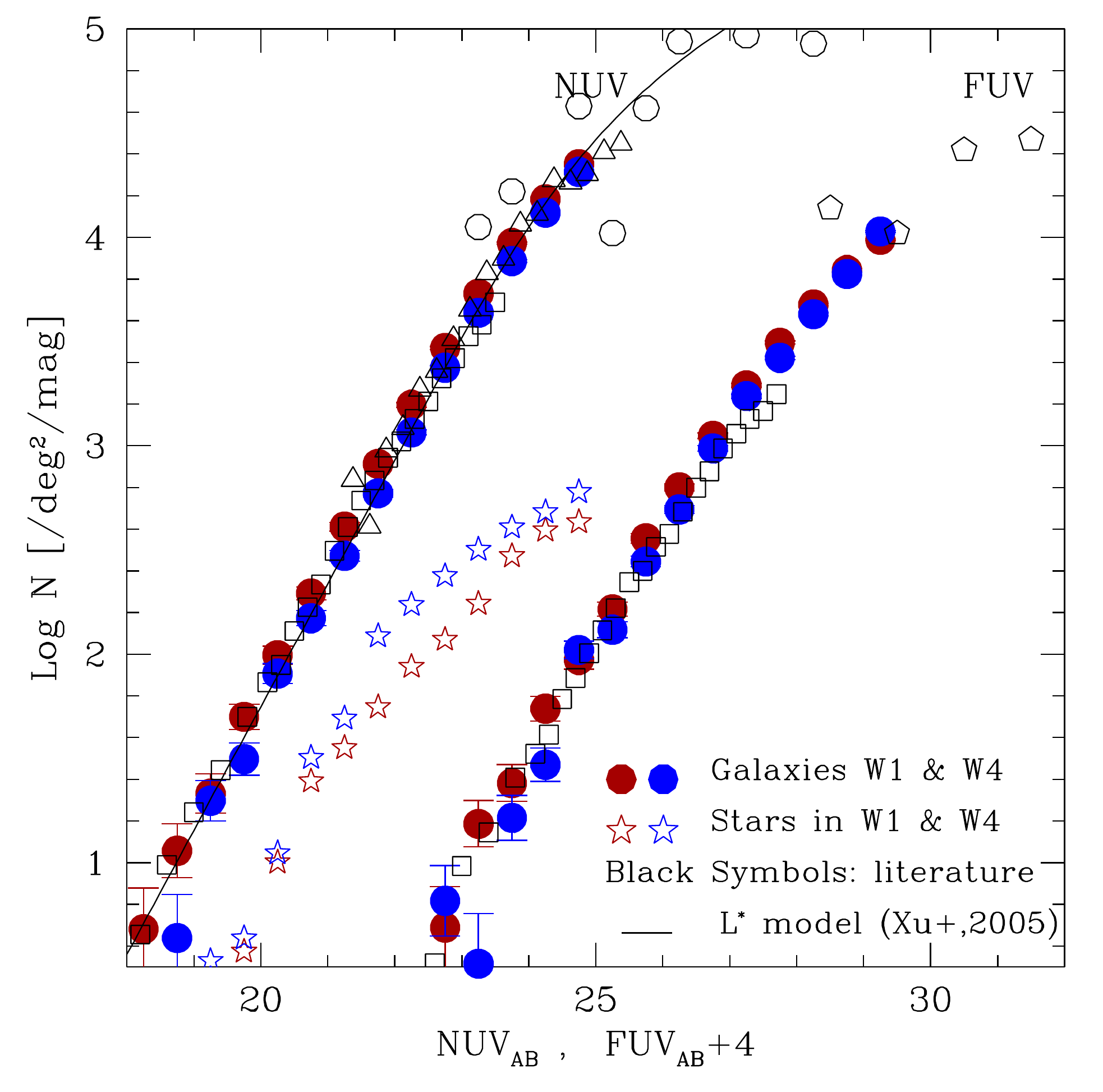}   
\caption[]{ GALEX number counts in W1 (red symbols) and W4 (blue symbols) regions for the galaxies (filled circles)  and the stellar population (open stars).  The FUV magnitudes are shifted by +4 mag.  Open black symbols are from the literature (triangles: \citet[]{Hoversten2009}; circles and pentagons: \citet[]{Gardner2000}; squares: \citet[]{Xu2005}  ) }
\label{fig:Nuv}
\end{figure}

 To conclude, the EMphot algorithm allows us to address the problem
that is related to the GALEX confusion. It provides a significative improvement with respect to the GALEX pipeline photometry for the deep fields and allows us to make full use of the GALEX photometry down to the nominal S/N of the observations.  The GALEX number counts with EM photometry in the FUV and NUV channels are shown in Fig.~\ref{fig:Nuv} separately for stars and galaxies (see Sect.~\ref{sec:sg}), and they are compared with the literature. 

%

\section{Classification of stars, QSOs, and galaxies}
\label{sec:sg}
\begin{figure}[!ht]
\centering
\includegraphics[width=0.9\hsize]{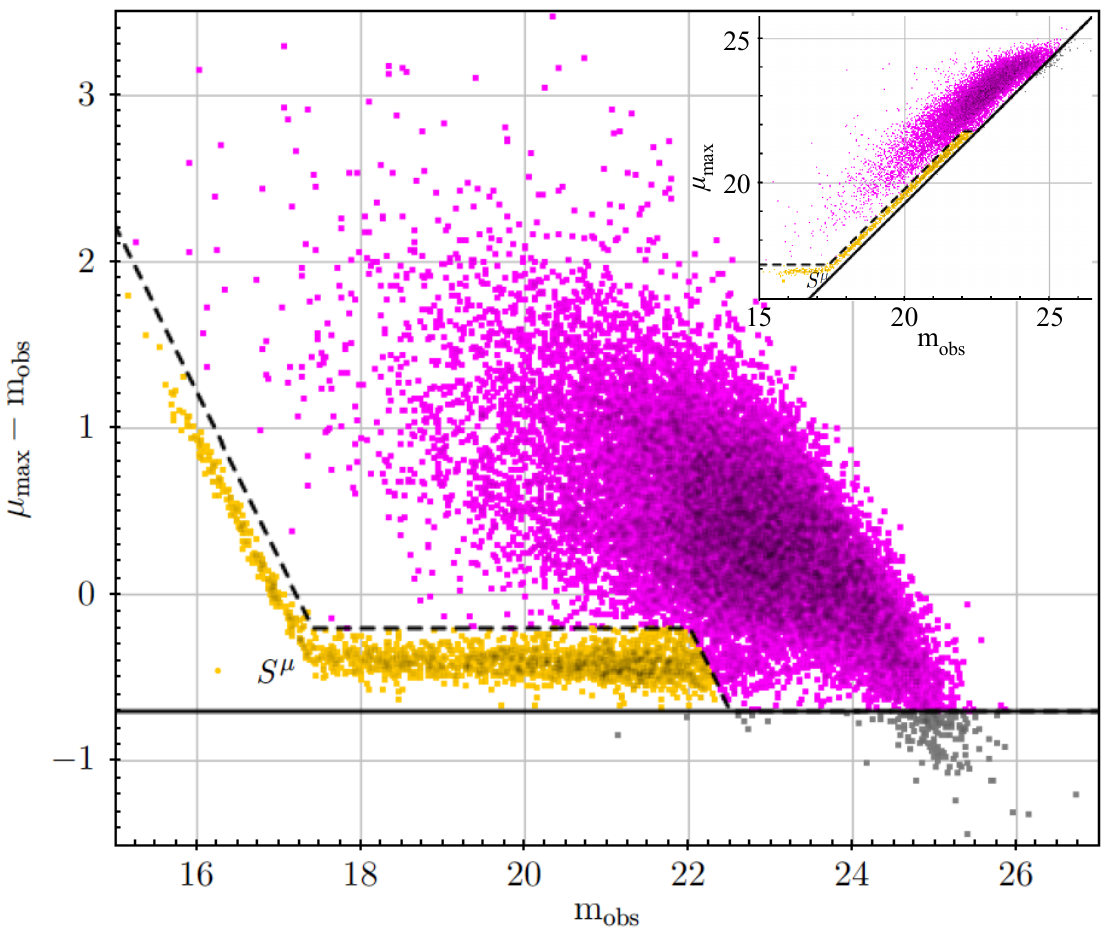}
\caption{Photometric classification between point-like and extended sources in the $\mu_{max}-m_{obs}$ vs $m_{obs}$  plane (for one tile of the CFHTLS in the $i$-band). The dashed line represents our selection of point-like sources ($\textit{\textbf{S}}^{\mu}$, yellow dots). All objects lying below the solid line (gray dots) are considered as false detections. 
\label{stargal_sel}}
\end{figure}

 To minimise the misclassification of stars and galaxies in the final samples, special care must be applied to the star-galaxy separation. For instance in the W4 field, near the Galactic plane,  the VVDS survey (based on a pure $i < 22.5$ target selection)  found more than 32\% of the spectroscopic sources to be stars \citep{Garilli2008}.
  To distinguish between stars and galaxies without damaging the completeness of our sample, we performed a classification based on three different diagnostics: \\
\begin{itemize}
\item[•] We used the maximum surface brightness versus magnitude (hereafter $\mu_{max}-m_{obs}$) plane where bright point-like sources are  clearly separated from galaxies \citep[see][]{Bardeau2005, Leauthaud2007}, as shown in Fig.~\ref{stargal_sel}. The method relies on the fact that the light distribution of a point-like source and its magnitude are proportional. Moreover, it allows us to detect objects with light distributions that are more peaked than the point-spread function (PSF), which are generally fake detections ($gray$ dots) and then removed. In the inset panel of Fig. \ref{stargal_sel}, one can see that the $\mu_{max}$ selection ($yellow$ dots), noted $\textit{\textbf{S}}^{\mu}$,  is only relevant up to a certain surface brightness limit, $\mu_{max}^{lim}$. The limit depends on the seeing, thus varies from tile to tile and with the considered passband. We performed the $\mu_{max}$ selection on the g,r,i passbands and imposed that a source is  classified as compact if it belongs to $\textit{\textbf{S}}^{\mu}$ in at least two bands;
\item[•] we used the SED fitting technique to compare the reduced $\chi^2$ obtained with the galaxy templates described in Sect. \ref{zp_method} and a representative stellar library from \citet{Pickles1998}. An object belongs to $\textit{\textbf{S}}^{\chi}$ if  $\chi^2_{star} / \chi^2_{gal} < 2$, where the factor 2 has been set to catch 90\% of the point-like objects;
\item[•] to minimise the loss of galaxies in the above steps, we finally used the $(g-z)$ vs $(z-K)$ colour diagram \citep[equivalent to the $BzK$ plane of][]{Daddi2004} when available to isolate the stellar sequence, noted $\textit{\textbf{S}}^{BzK}$. 
\end{itemize}  
We adopted the latter as a sine qua non condition to be a star, which leads to these final combinations for a source to be classified as star:
\begin{eqnarray}
[ \textit{\textbf{S}}^{\mu} \cup \textit{\textbf{S}}^{\chi} ] \cap \textit{\textbf{S}}^{BzK}  ~~~ \mathrm{if} ~~\mu_{max} < \mu_{max}^{lim} ~;  \nonumber \\
\textit{\textbf{S}}^{\chi} \cap \textit{\textbf{S}}^{BzK}   ~~~  \mathrm{if} ~~\mu_{max} > \mu_{max}^{lim}  ~.
\label{eq_sel}
\end{eqnarray}

In the bright point-like source domain ($\mu_{max} < \mu_{max}^{lim} $), we define as QSO the objects lying on the galaxy side of the $BzK$ diagram 
(corresponding to $\textit{\textbf{S}}^{\mu}\ \  \cap\ \  ! \textit{\textbf{S}}^{BzK}$). Dominated by their nucleus, the emission of those AGNs is currently poorly linked to their stellar mass. They represent less than 0.5\% of the global sample.
By using 1119 spectroscopically confirmed AGNs from the SDSS and VIPERS surveys, we confirmed that our photometric selection of QSOs is able to catch $65\%$ of them, without losing a single galaxy.

All the objects that were not defined as stars or QSO were then considered as galaxies. We verified on a sample of 1241 spectroscopically confirmed stars that we thereby catch 97\% of them, while we keep more than 99\% of our spectroscopic galaxy sample. With this selection we finally found and removed $\sim$ 8\% and $\sim$ 19\% of stars at $K_s$ < 22 for W1 and W4, respectively (in the unmasked area). The star number counts in the WIRCam and GALEX (FUV, NUV) passbands are shown separately for the W1 and W4 fields in Figs.~\ref{fig:Nk} and~\ref{fig:Nuv}, respectively. The density of stars in W4 is higher than in W1 at all wavelengths and reaches almost a factor 3 in $K_s$ band.  

\begin{figure}[!ht]
\centering
\includegraphics[width=0.9\hsize]{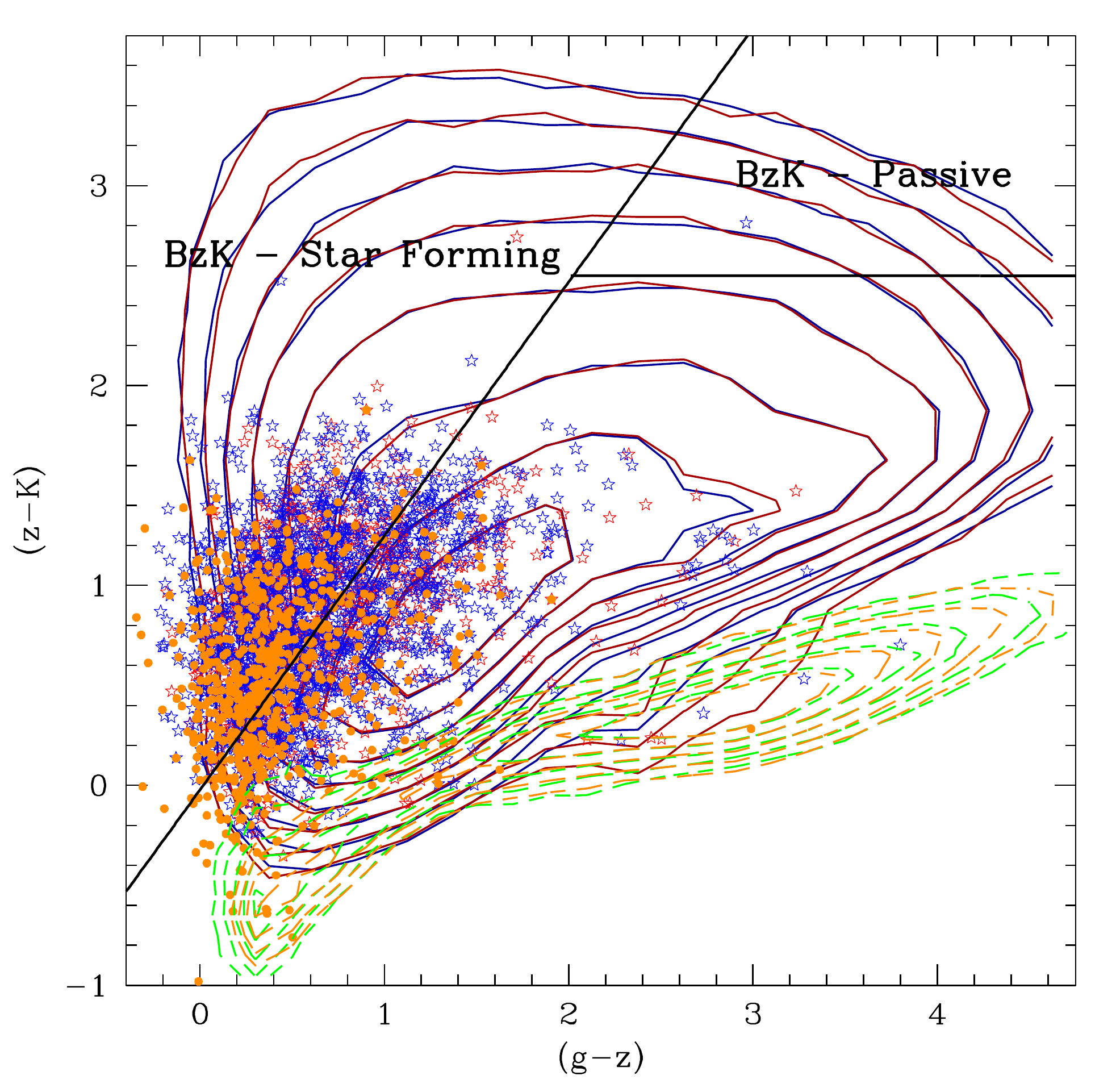}
\caption{Distribution of galaxies, stars, and QSOs in the $(z-K)$ vs $(g-z)$ colour diagram. Density levels are in logarithmic scale ($\Delta$=0.3dex) for galaxies (solid lines in W1 and W4 fields shown blue and red, respectively) and stars (dashed lines in green and orange). Objects from our QSO samples are shown as open stars (blue: W1; red: W4), while the SDSS/VIPERS AGN sample is shown as orange dots. The solid black lines delineate the high-redshift BzK populations (star-forming and passive galaxies).   
  \label{fig:bzk_plot}}
\end{figure}

Finally, in Fig.~\ref{fig:bzk_plot} we show the star and galaxy distributions in the $BzK$ colour diagram for the two fields separately. The low-density contours of the galaxy distributions can overlap the stellar locus. Our photometric sample of QSOs (red and blue stars) is also compared with spectroscopically confirmed AGNs (orange dots) from the SDSS and VIPERS surveys.

%

\section{Photometric redshifts }
\label{zp_method}
\subsection{Photometric redshift measurement}
Thanks to the enlarged wavelength coverage, we can estimate new photometric redshifts for the T0007 CFHTLS sources in the VIPERS regions,  which also benefit from the large spectroscopic coverage for calibration purposes.  The photometric redshifts were computed with the SED-fitting code \lephare \citep{Arnouts2002,  Ilbert2006}. We adopted the galaxy SED templates used by
\citet{Ilbert2009} (31 empirical templates from \citet{Polletta2007}
and 12 star-forming templates from Bruzual and Charlot models
\citep[]{BC2003}). They were modified to better fit the
photometry discussed here. We used the large VIPERS spectroscopic
redshift sample \citep[see][for details]{Coupon2015}.  The extinction
was added as a free parameter with a reddening excess E(B-V) < 0.3
following different laws: \citet{Prevot1984}, \citet{Calzetti2000},
plus a dusty Calzetti curve including a bump at 2175\AA. No extinction
was allowed for SEDs redder than Sc. The \citet{Prevot1984} extinction
law was allowed for templates redder than SB3 templates
\citep[see][]{Ilbert2009} and \citet{Calzetti2000} for bluer ones.
Finally, the systematic disagreement between the photometry and the
template library was corrected by \lephare according to the
method described by \citet{Ilbert2006}. In brief, the code tracks a systematic shift between the predicted  and the observed magnitudes in each band at known redshifts. Since our observation area is divided into 54 CFHTLS tiles of $\lesssim 1 deg^2$ with photometry varying from tile to tile, we performed a tile-by-tile colour optimisation. 
We used the median offset over all the tiles when there were not enough
available galaxies with spectroscopic redshift in the tile
($N_{gal}^{spec} \leq 100$), which was the case in 12 tiles.

\subsection{Comparison with the spectroscopic sample}

Our WIRCam survey has been designed to cover the VIMOS Public Extragalactic Survey \citep[VIPERS;][]{Guzzo2014} carried out with the VIMOS\footnote{Please refer to \citet{LeFevre2003} for more details.} spectrograph. 
We used the first public data release (PDR1) described in \citet{Garilli2014}, which provides a $i < 22.5$-limited sample of $\sim57,000$ spectroscopic redshifts at $0.5 < z < 1.2$ \footnote{The targets were preselected in colour space to enhance the probability of observing galaxies at $z > 0.5$. This technique results in $\sim 40\%$ sampling rate with a single pass of VIMOS.} with high precision ($\sigma_v=140 km s^{-1}$ , which corresponds to $0.0005 (1 + z)$ ).

 The spectroscopic sample was complemented with other surveys listed in Table \ref{tab_zs_info}. We outline their characteristics
in this table.  We used the VIMOS Very Deep Survey
\citep[VVDS,][]{LeFevre2013} "Wide" ($i$ < 22.5) in W4 and "Deep" ($i$
< 24) in W1, spanning the redshift ranges $0.05 < z < 2$ and $0.05 < z
< 5$ respectively. Only a fraction of the VVDS survey intersects our
WIRCam coverage. 
 The deep part of the VVDS survey allowed us to test the photometric redshifts in a fainter regime and at higher redshift. For the same purpose, we also included the $K < 23$ limited UKIDSS spectroscopic Ultra Deep Survey \citep[UDSz,][]{Bradshaw2013,McLure2013}, which
has redshifts in $0 < z < 4.8$ obtained with the VIMOS and FORS2 VLT spectrographs. 
 We used redshifts from the Baryon Oscillation Spectroscopic Survey \citep[BOSS,][]{Dawson2013} when available. BOSS is a bright-limited ($i < 19.9$) spectroscopic survey from the Sloan Digital Sky Survey (SDSS), providing redshifts up to $z = 0.7$ in the entire region of our $K_s$-coverage.  
  Finally, we included the most secure redshifts from the PRIsm MUlti-object Survey \citep[PRIMUS;][]{Coil2011}. This sample is based on low-resolution spectra ($\lambda/\Delta \lambda \sim 40$) and is limited to $i \sim 23$ \citep[see][for more details]{Cool2013}. 

We selected only the most secure spectroscopic redshifts, which means confidence levels above 95\% for high-resolution surveys: 
\begin{itemize}
\item  flags 3 and 4 with VIMOS (VIPERS, VVDS, UDSz) 
\item  flags A and B with FORS2 (UDSz)
\item  z-warning 0 in BOSS 
\end{itemize}
and $\sigma < 0.005$ (8\% of outliers with $\delta z/(1+z) > 5 \sigma)$ for PRIMUS flag-4 redshifts. When they were available, the redshift measurements from VIPERS were used. Otherwise, the measurements from the deepest high-resolution spectra were favoured.

In total, we collected a sample of $\sim 51396$ high-quality spectroscopic redshifts to calibrate and measure the accuracy of our photometric redshifts over the entire area of the survey.

\begin{table}
  \vspace{0.3cm}
  \caption{Characteristics of the spectroscopic redshift sample. We report the number of spectroscopic redshifts ($N_{gal}^{spec}$), the median value in the $i$-band ($i_{med}$), the contribution of each survey to the total spectroscopic sample ($\mathcal{F}_\%$), and their relative fraction of galaxies with $i < 22.5$ and $i>22.5$.
 \label{tab_zs_info}  }
  \centering
  \begin{tabular}{l*{4}{c}}
  \hline
  \noalign{\medskip}
  Spectroscopic & \multirow{2}{*}{$N_{gal}^{spec}$}  & \multirow{2}{*}{ $i_{med}$} &  \multirow{2}{*}{\begin{large} ~~$\mathcal{F}_\%$ \end{large}$^{~i<\textbf{22.5}}_{~i>22.5}$ }   \\ 
  survey &  &  &  \\ 
  \hline
  \hline
  \noalign{\medskip}
  VIPERS  & 33896 & 21.7 & 60.5$~ ^{\textbf{95.7}}_{4.3}$ \\
  \noalign{\smallskip}
  PRIMUS & 14725 & 21.7 & 26.3$~ ^{\textbf{78.3}}_{21.7}$ \\
  \noalign{\smallskip}
  SDSS-BOSS & 4075 & 19.0 & 7.3$~ ^{\textbf{100.0}}_{0.0}$ \\
  \noalign{\smallskip}
  VVDS Wide & 1630 & 21.6 & 2.9$~ ^{\textbf{96.5}}_{3.5}$ \\
  \noalign{\smallskip}
  VVDS Deep & 457 & 23.1 & 0.8$~ ^{\textbf{25.0}}_{75.0}$ \\
  \noalign{\smallskip}
  UDSz & 1220 & 23.1 & 2.2$~ ^{\textbf{27.3}}_{72.7}$ \\
  \noalign{\smallskip}
  \hline
 \noalign{\smallskip}
  Total sample & 51396 & 21.6 & 100$~ ^{\textbf{89.4}}_{10.6}$ \\
  \noalign{\smallskip}
  \hline
  \noalign{\smallskip}
\end{tabular}
\end{table}

\begin{figure}
\includegraphics[width=8.5cm,height=\hsize]{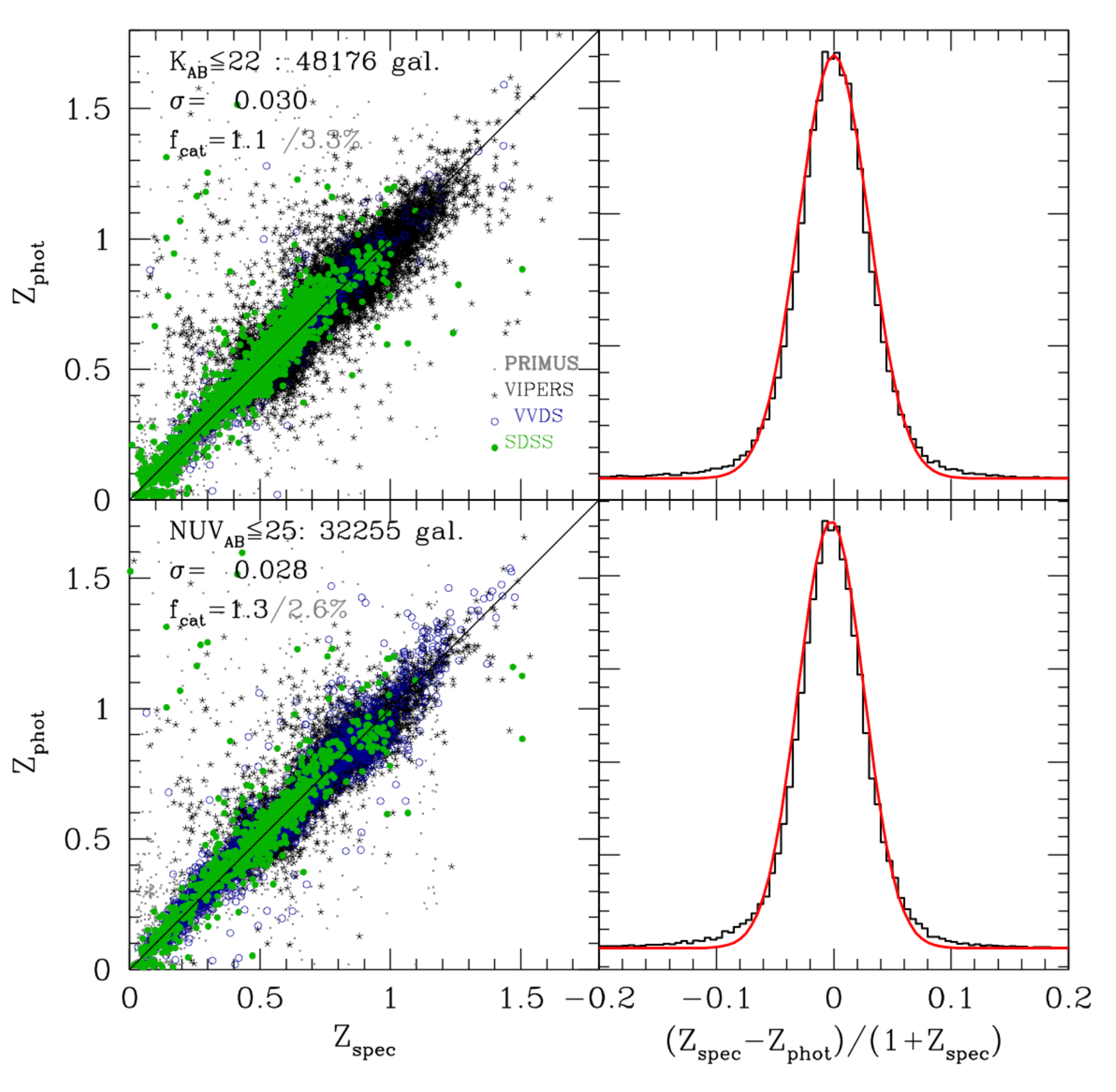}   
\caption[]{Comparison of our photometric redshifts with the VVDS, VIPERS, PRIMUS, and BOSS most secure spectroscopic redshifts, for  galaxies with  $K\le 22$ (top panels) and $NUV\le 25$ (bottom panels). We report the dispersion ($\sigma_{NMAD}$) by combining all the surveys, while the catastrophic fractions ($f_{cat}$ defined as $|\Delta z|/(1+z_s) >0.15$) are given for all surveys but PRIMUS (black value), which is shown separately (gray value). The observed distributions of photometric errors are compared to a Gaussian distribution with the same $\sigma$ (right panels).  }
\label{fig:zphot}
\end{figure}

\begin{figure}
\includegraphics[width=8.5cm,height=\hsize]{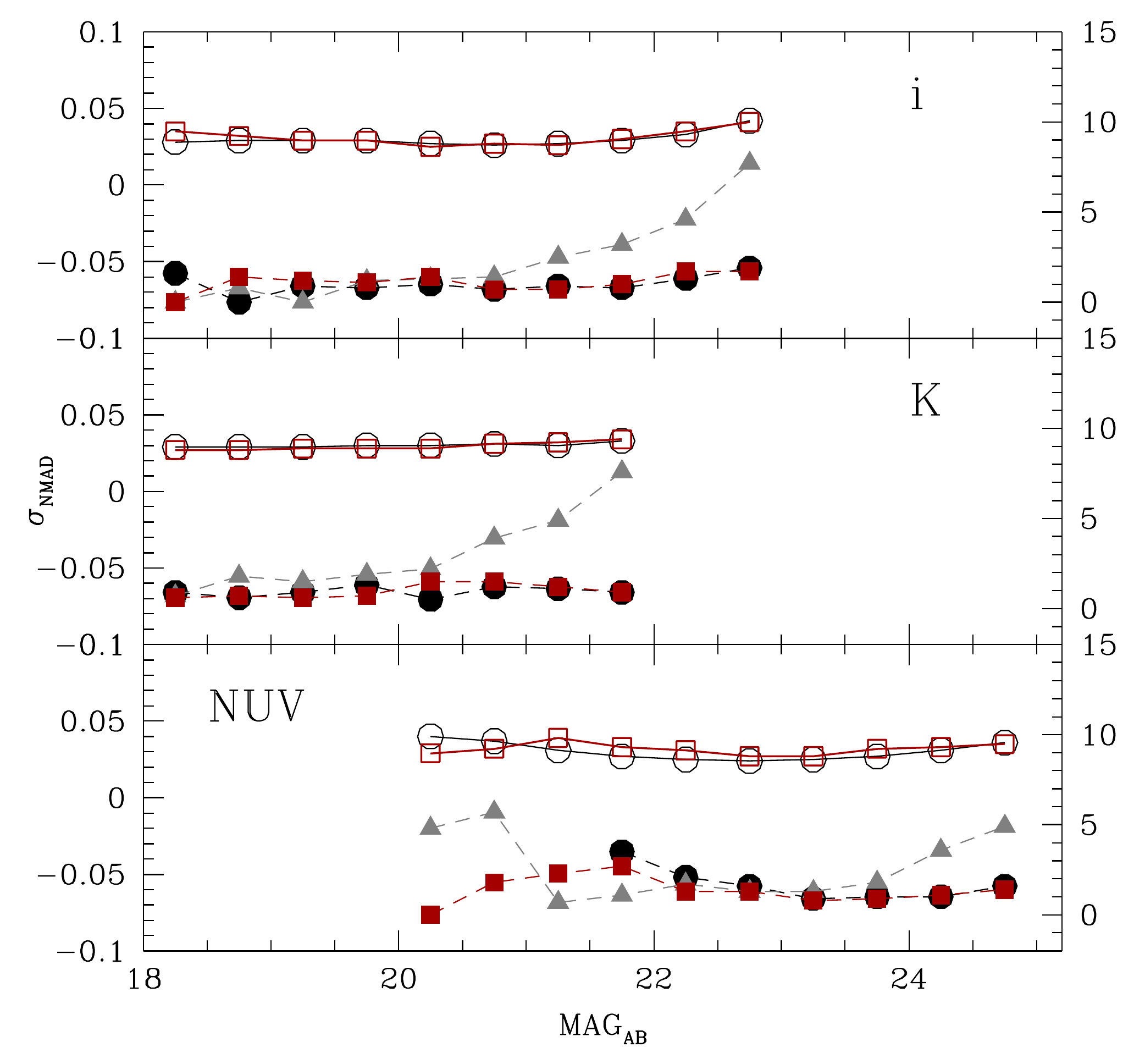}   
\caption[]{Comparison of the dispersion ($\sigma_{NMAD}$) as a
  function of magnitude (top panel: $i$; middle panel: $K_s$; bottom
  panel: NUV). The $\sigma_{NMAD}$ for the W1 (black circles) and W4
  (red squares) fields correspond to the open symbols and solid lines
  and refer to the left y-axis.  The fraction of catastrophic
  redshifts are reported as field symbols and dashed lines and refer
  to the right y-axis (\%).  The PRIMUS spectroscopic sample in W1
  field is shown separately (gray triangles).  }
\label{fig:dz_sel}
\end{figure}

Figure \ref{fig:zphot} shows the comparisons between our
photometric redshifts and the spectroscopic redshifts for the $K\le
22$ (top panels) and $NUV\le 25$ (bottom panels) selected samples.  
 By using the NMAD to define the scatter\footnote{$\sigma_{\Delta z/(1+z)} = 1.48~median(~|z_{spec}-z_{phot}| / (1+z_{spec})~)$}, we
find $\sigma_{\Delta z/(1+z)}\lesssim 0.03$ for both samples with an
outlier rate\footnote{$\eta$ is the percentage of galaxies with
$\Delta z/(1+z)>0.15$} $\eta\sim 1\%$.  
The photometric redshift errors are well described by a Gaussian distribution with the
same $\sigma$ for the two samples, as illustrated in the right panels of
Fig. \ref{fig:zphot}. 
As reported in Table~\ref{tab_zs_info}, we recall that this comparison is dominated by spectroscopic galaxies with $i \le 22.5$. 
In Fig. \ref{fig:dz_sel} we show the $\sigma$ (left axis, open symbols
and solid lines) and the catastrophic fraction rate (right axis, solid
symbols and dashed lines) as a function of magnitude ($i$ : top panel;
$K_s$: middle panel; $NUV$: bottom panel) for the W1 and W4 fields.  
As expected, a slight deterioration of the accuracy at $i > 22$
is visible.

It is interesting to note that when the PRIMUS
redshifts are included in our statistics, we observe a strong increase in the average number
of catastrophic failures (by a factor 2, 3) in the $NUV-$ and $K_s$
-selected samples (as reported in Fig. \ref{fig:zphot}, gray values).
We show separately the catastrophic fraction for PRIMUS spectra in Fig. \ref{fig:dz_sel} 
(gray filled triangles). A strong increase is observed toward the faint magnitudes,
while we do not observe this behaviour for the high-resolution spectroscopic 
surveys (SDSS, VVDS, VIPERS, and UDSz). 
This could suggest that the PRIMUS redshift flags are too
optimistic regarding the quality of the spectra and redshift
measurements at fainter magnitudes\footnote{We recall that the average redshift success of PRIMUS declines to 45\% at $i \sim 22.5$ \citep[see][]{Cool2013}.}.  

Nevertheless, the large amount of high-resolution spectroscopic redshifts available in our area allows us to reliably quantify the quality of our photometric
redshifts. As shown in Fig. \ref{fig:dz_sel}, the dispersion is always below $\sigma \sim 0.04$ and the catastrophic fraction (excluding PRIMUS) always below 2\% over the entire magnitude range at $i \le 23$ (both down to $K_s \sim 22$ and $NUV \sim 25$). 
However, we recall that values quoted in Figs.~\ref{fig:zphot} and ~\ref{fig:dz_sel} do not account for the selection bias introduced by the spectroscopic sample against blue or red galaxies when moving toward fainter NUV or K magnitudes, respectively. 
 
%

 %
\subsection{Comparison with the VIDEO photometric redshifts}

Half of the three VISTA pointings from the VIDEO survey \citep[]{Jarvis2013} overlaps with our WIRCam observations (see Fig.~\ref{fig:iq}).  We can compare the photometric redshifts derived from the deep VIDEO Y, J, H, K data (reaching K $\sim$ 23.5) with our WIRCam photometric redshifts to partially remove the spectroscopic bias discussed above.
  
We first matched the full CFHTLS-T0007 optical catalogue with the NIR VIDEO photometry (within a search radius of $0.5\arcsec$),
to which we added the GALEX photometry.  We derived the photometric redshifts for this combined catalogue (GALEX-CFHTLS-VIDEO) using the same recipe as described above.

\begin{figure} 
\includegraphics[width=0.99\hsize]{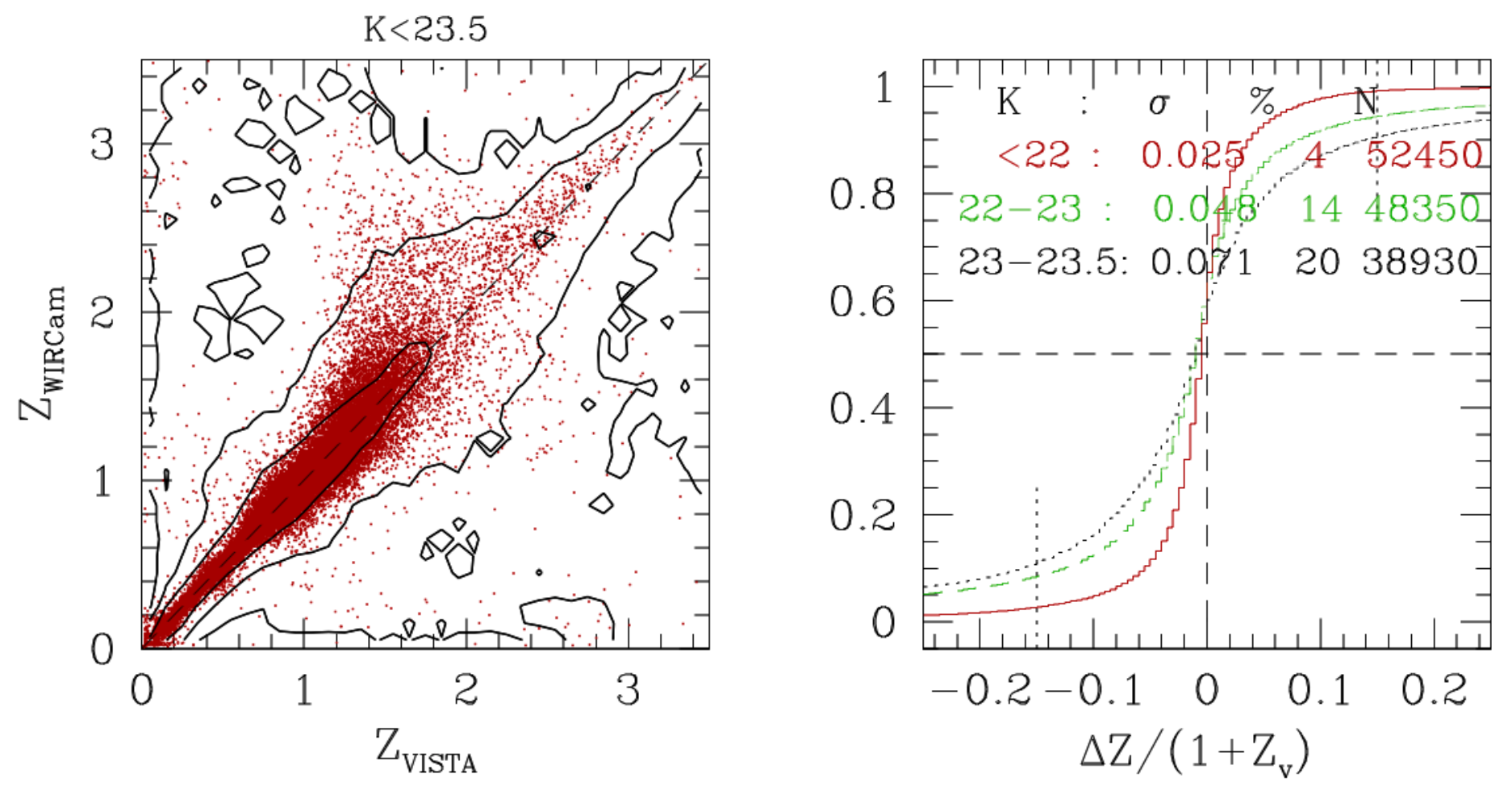}  
\caption[]{Comparison of the photometric redshifts measured using either WIRCam K band ($z_{\textsc{wirc}\small{am}}$) or  YJHK  photometry from VISTA ($z_{\textsc{vista}}$) for galaxies with $K_{VISTA}\le 23.5$. The left panel shows the comparison $z_{\textsc{wirc}\small{am}}$ vs $z_{\textsc{vista}}$ for $K_{VISTA}<22$ (red dots), while  the  density contours spaced in logarithmic scale of 1dex are for the faint sample, $22<K_{VISTA}<23.5$. The right panel shows the cumulative distributions of the difference ($\Delta z/(1+z_{\textsc{vista}})$) split into three K-band subsamples ($K_{VISTA}\le 22$: red line; $22<K_{VISTA}<23$: green line; $23<K_{VISTA}<23.5$: red line). We also report the dispersions ($\sigma$), the fractions of catastrophic sources (in \%), and the number of sources used.
}
\label{fig:dzv_k}
\end{figure}

\begin{figure}
\includegraphics[width=\hsize]{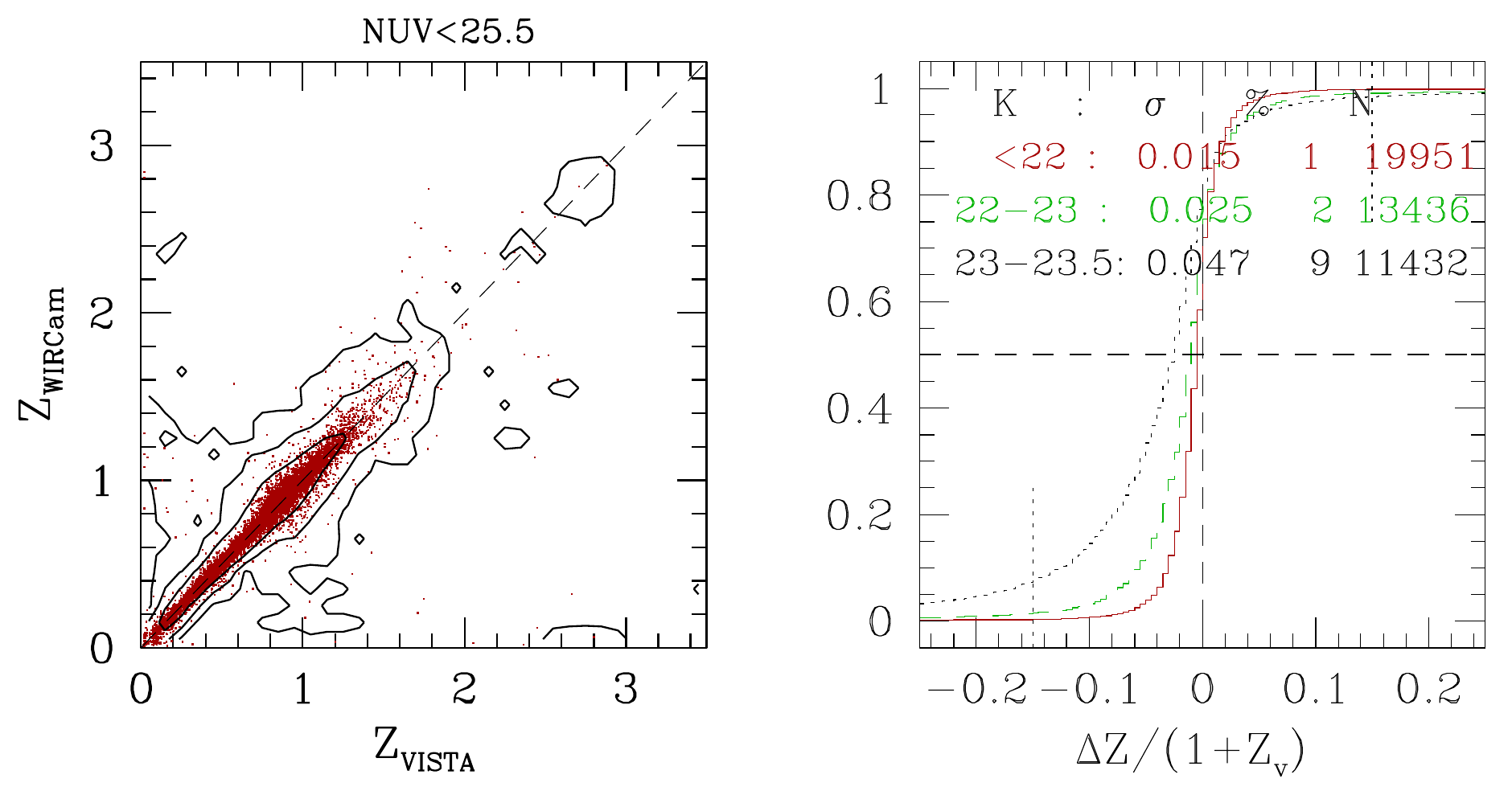}   
\caption[]{Same as Fig.~\ref{fig:dzv_k} for a NUV-selected sample with $NUV\le 25.5$ and also split into  $K_{VISTA}$  subsamples. 
}
\label{fig:dzv_nuv}
\end{figure}

In Figs. \ref{fig:dzv_k} and \ref{fig:dzv_nuv} we compare our
original photometric redshifts (referred to as $z_{\textsc{wirc}\small{am}}$) with 
VIDEO photometric redshifts (referred to as $z_{\textsc{vista}}$) for a K- ($K_{VISTA}\le 23.5$) and NUV- ($NUV\le 25.5$) selected sample.  In both figures we split the data into a K-bright ($K_{VISTA}\le 22$) and a K-faint ($22\le K_{VISTA}\le 23.5$) sample.   In the left panels, we show the one-to-one comparison of the photometric redshifts for the K-bright samples (red dots), while the K-faint samples are shown with density contours spaced in logarithmic scale of 1dex (black lines). In the right panels we show the cumulative distributions of the errors ($\Delta z/(1+z)$), and  we report the number of sources in each selection, the dispersions, and the catastrophic fractions.
 
For the $K$ -selected sample in Fig.~\ref{fig:dzv_k}, our photometric redshifts are consistent with $z_{\textsc{vista}}$  ($\sigma \sim 0.025$, $\eta\sim$ 4\%) up to $z\sim 3$ at bright
magnitudes. 
The $z_{\textsc{wirc}\small{am}}$ tend to overestimate the redshifts in between $1<z_{\textsc{vista}}<2$.  
In the two fainter magnitude bins, the photo-z remains unbiased up to $z\sim 3,$ but the dispersions and the catastrophic fractions quickly increase. This is expected since our photometric redshifts are only constrained by the optical magnitudes of the CFHTLS.
  
For the NUV selected sample in Fig.~\ref{fig:dzv_nuv}, the
photometric redshifts are in excellent agreement up to redshift $z\sim
1.5$. For the bright subsample, $K_{VISTA}\le 22$ (red dots), 
we observe a very small dispersion ($\sigma\sim 0.015$) and a negligible fraction of outliers. This accuracy holds down to the magnitude range $22\le K_{VISTA}\le 23$.  At fainter magnitudes, $K_{VISTA}\ge 23$,   our photometric redshifts starts to become slightly worse and has a small bias $\Delta z/(1+z) \le 0.02$. 
 This shows that for our NUV-selected samples, we have a good confidence level in the photometric redshifts derived with WIRCam. 

%
   
%
%
%
%
\section{Discussion}

\subsection{BzK selection}
\begin{figure}
\includegraphics[width=9.2cm,height=1\hsize]{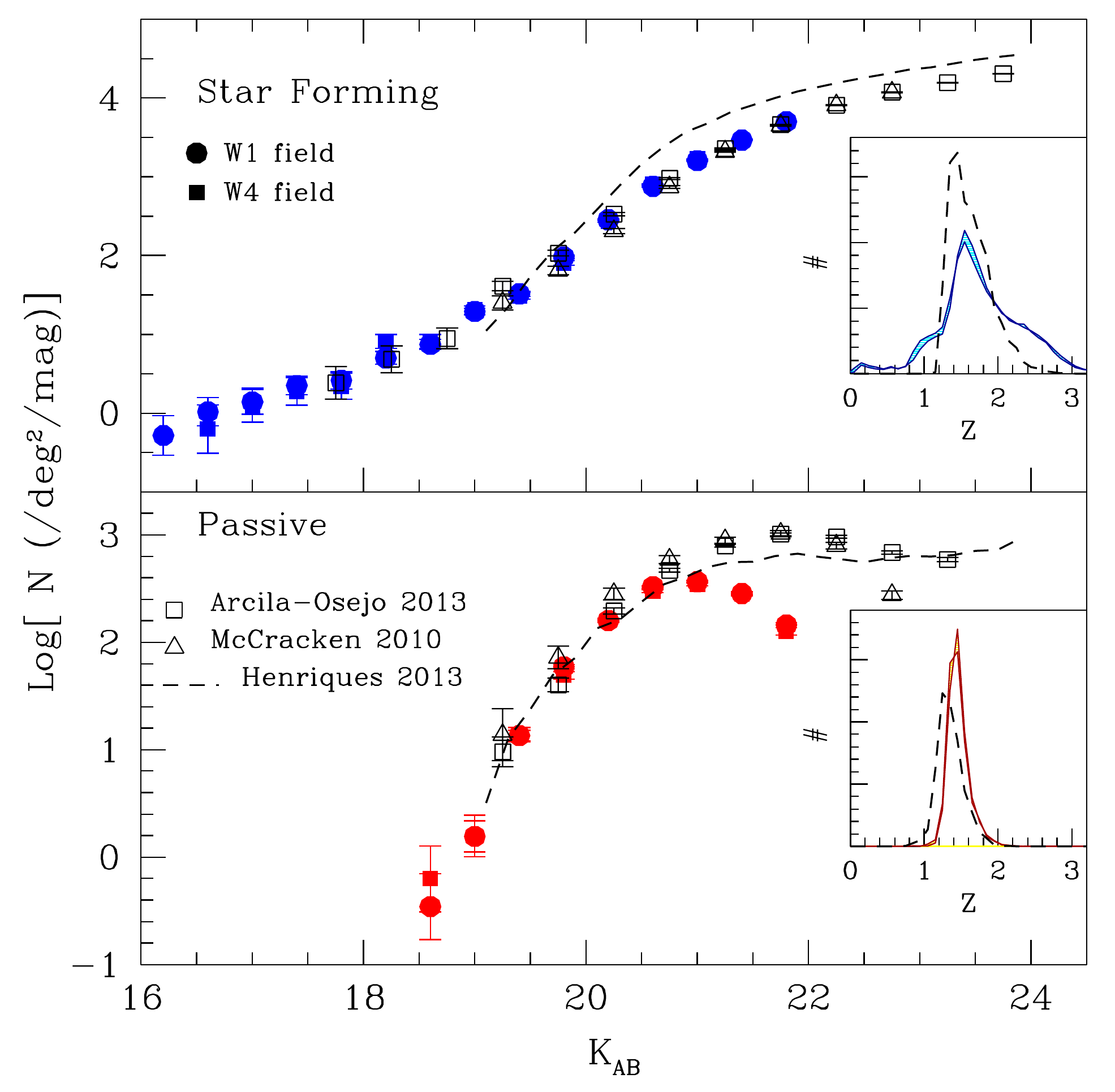}   
\caption[]{Differential number counts of star-forming (top) and passive (bottom) BzK galaxies in the W1 and W4 fields. We compare with deeper observations in CFHTLS deep fields \citep[open squares,][]{Arcila2013} and COSMOS \citep[open triangles,][]{McCracken2010}  and model predictions \citep[dashed lines,][]{Henriques2013}  The inset panels show the photometric redshift distributions with $K_{AB}\le 22$. }
\label{fig:bzk}
\end{figure}

The large area covered by our K-band survey and the optical depth of
the CFHTLS Wide data makes our sample unique to select a large sample
of bright and massive galaxies at high redshift, $z\ge 1.4$, with the
$BzK$ colour selection \citep[]{Daddi2004}. We adopted the criteria from
\citet[]{Arcila2013}, who adjusted the selection for the CFHTLS and
WIRCam K-band photometry. This colour selection is shown in
Fig.~\ref{fig:bzk_plot} (solid lines) and provides a sample of
$\sim$100,000 star-forming and $\sim$7,800 passive BzKs. The
differential number counts are shown in Fig. \ref{fig:bzk} for the
star-forming (top panel) and passive (bottom panel) galaxies down to $K\sim
22$.  The error bars are from Poisson statistics. Our results are
compared with the deep near-IR data in the four CFHTLS deep fields
\citep[only the average of the four fields is shown, ][]{Arcila2013}
and in the COSMOS field \citep[]{McCracken2010}. Our results provide
robust constraints at the bright end. For the star-forming BzKs, we
agree well with previous works in the overlapping
range and down to our limit, $K_s\sim$22.  For the passive BzKs our
result agrees with others at the bright end, but we observe a drop at
$K_s\ge20.7$.  Because of the optical $gri$ $\chi^2$ image detection,
our data suffer from incompleteness against the reddest
population. As illustrated in Fig.~\ref{fig:kcomp}, the 50\%
completeness limit corresponds to $(z-K_s)\sim 3$ at $K_s\sim$21 and
drops to $(z-K_s)\sim 2.5$ at $K_s\sim$22 (corresponding to the passive $BzK$ colour threshold). At high redshift, $z\ge 1.4$, our native optical selection introduced
 a bias against those passive galaxies, which explains the observed drop in density of 
 the passive BzK population at $K_s\ge 20.7$.
 
In the insets of each panel of Fig.~\ref{fig:bzk}, we show the
photometric redshift distributions for the two populations in the two
fields.  The vast majority of BzKs is observed at $z_{phot}\ge 1.2$,
as expected from the original work by \citet[]{Daddi2004}, and with a
negligible fraction of outliers at low redshift. The star-forming BzKs
 show a large spread in redshift, $1\le z\le 3$, with a
peak around $z\sim 1.5$. Passive BzKs also peak at $z\sim 1.5,$ but
show a narrower distribution ($1.2\le z\le 1.8$).
   
We compared the observations with the semi-analytical model of
\citet[]{Henriques2013}, based on the Millenium simulations, which
has been tuned to reproduce the stellar mass functions at all redshifts. We defined our own colour criteria in the simulations  to best distinguish the galaxies with high and low specific SFR (sSFR) at $z\ge 1.4$. We adopted the following criterion: $(z-K)-0.9(g-z)\ge 0.7$ for active and $(z-K)-0.9(g-z)\ge 0.7$ \& $(z-K)\ge 2.55$ for passive galaxies. The predicted number densities as a function of magnitude are shown in Fig.~\ref{fig:bzk}.
The SAM simulation shows similar behaviours as the observed ones for both the star-forming and the passive galaxy populations. The passive population agrees well over the whole magnitude range. For the  star-forming population, the simulation over-predicts the density by at least a factor two at magnitudes fainter than $K_{AB}\sim20$. This is highly significant in view of the uncertainties in the observations. The redshift distributions in the simulation are concentrated in the redshift range $1\le z\le 2,$ and the redshift peaks are at slightly lower redshifts than our BzK photo-zs for both the active and the passive BzK populations. This difference can be partially due to the photo-z bias affecting this redshift range, which are spread toward higher redshift as discussed in Sect. 4.3.

\subsection{Galaxy morphology in the NUVrK diagram}
 The  star-forming and quiescent galaxies form a bimodal distribution in rest-frame colours or D4000 break. However, the separation between the two populations is not straightforward because of the presence of dust obscured star-forming galaxies when a single colour diagnostic is used. \citet[]{Williams2010} broke this degeneracy by using a bi-colour $(U-V)$ vs $(V-J)$ diagram ($UVJ$), which better separates the passive red galaxies from the dusty red galaxies. \citet[]{Arnouts2013} adopted a similar diagram $(NUV-r)$ vs $(r-K)$,  ($NUVrK$) to optimize the role of the current SFR and the dust obscuration.
   
\begin{figure*}
\centering
\includegraphics[angle=90,width=\hsize]{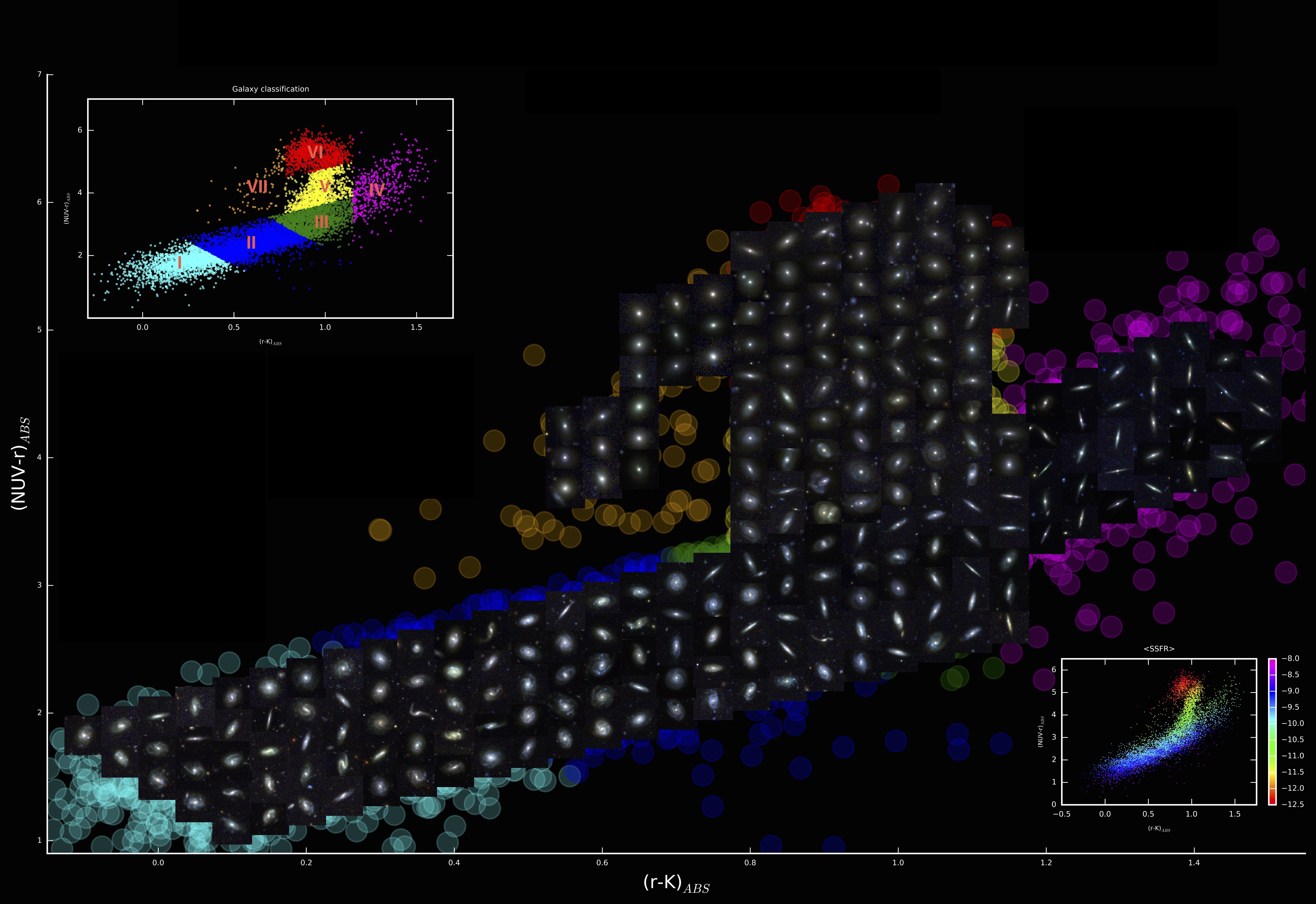}   
\caption[]{Illustration of the galaxy morphology in the $NUVrK$ diagram for galaxies with $z\le 0.25$.  Galaxies are split into seven categories shown in the top left inset, and background enlarged points are colour-coded according to their categories. Only the largest galaxies (with semi-major axis $A>50$ pixels) are shown. The lower right inset shows the behaviour of the sSFR (log-scale colour-coded).  }
\label{fig:morph}
\end{figure*}

 In this section we use the high image quality of the CFHTLS and the large volume to investigate the morphological appearance of well-resolved low-z galaxies according to their location in the $NUVrK$ diagram.  The full sample of galaxies with $z\le 0.25$ was split into seven broad categories shown by different colours in the top left panel of Fig.~\ref{fig:morph},  while the sSFR is colour-coded in the bottom right panel. 
In the central part of Fig.~\ref{fig:morph}, we show the a set of $gri$ colour stamps \citep[generated with the STIFF software;][]{Bertin2011} of the largest galaxies (with a major axis $A\ge 50$ pixels). They are randomly placed within subregions (splitting each category into four subcategories).

 We are indeed limited in the number of galaxies that we can show in this diagram, and we favoured the most attractive galaxies. However, we generated several realisations of the diagram and found that it does not change the points we address below on the morphological appearances within the $NUVrK$ diagram:
\begin{itemize}
\item Star-forming galaxies fall in the categories I to IV. The bluest galaxies (category I$-$II)  have spiral or irregular morphologies and often show clumps of star formation and more disturbed disks. They have high specific SFRs 
 that can reflect a high gas fraction. For the oldest population (redder $(r-K)$ colour, category II$-$III)  we observe a larger portion of grand-design spiral galaxies. They are more massive and have a lower sSFR. Their morphologies suggest that these spiral galaxies are more settled and more stable than their lowest mass counterparts, which is consistent with the kinematic analysis of spiral galaxies from \citet[]{Kassin2012} .
\item  Category IV (purple) reveals a specific population of edge-on galaxies. This confirms the claim of \citet[]{Arnouts2013} that only highly inclined disk galaxies can produce such extreme infrared excess (IRX) and red $(r-K)$ colour \citep[see also][for a similar conclusion from UVJ diagram]{Patel2011}.   
\item Category V (yellow) shows a variety of morphologies. According to their decreasing sSFR, it represents massive galaxies most probably transiting from the star-forming to the quiescent populations.  We observe a significant fraction of mergers and post-mergers with extended tidal debris, consistent with the scenario of \citet[]{Hopkins2006},  where mergers trigger both the quenching through AGN feedback and the morphological transformation. 
Conversely, some of these interaction or merger systems can also be the result of gas accreted onto red galaxies from a minor gas-rich merger, which triggers new star formation activity. This rejuvenation process is observed in the local Universe \citep[]{Salim2010, Thomas2010} and also predicted in hydrodynamical simulations \citep[]{Trayford2016}. 
We also observe several spiral galaxies with bars and rings within the disks.  This could point us toward the importance of secular evolution in the star formation regulation and the disk instability leading to a morphological change.  \citet[]{Sheth2008} showed that most massive spiral galaxies have a higher fraction of bars, associated with a bulge and have redder colours, consistent with this result. 
\item Category VI (red)  is exclusively dominated by massive S0 and elliptical 
morphologies \citep[e.g.,][]{Bell2008}.  At  $z\sim 1$,  \citet[]{Ilbert2010} observed a significant fraction of quiescent galaxies with disk-dominated morphologies \citep[see also][]{Bundy2010}, but such a population is not observed in our local sample. 
\item Category VII (orange) shows a small population of galaxies with blue $(r-K)$ colour and $(NUV-r)$ redder than the star-forming categories. In the companion paper \citep{Moutard2016b}, we interpret this region through the quenching channel that characterises \textit{young} (blue $(r-K)$ colour) low-mass galaxies.  Evolutionary tracks in the $NUVrK$ diagram showed that their quenching should be faster than 0.5 Gyr.  Environmental quenching  such as harassment or ram-pressure stripping can be consistent with this timescale, but we exclusively observe elliptical morphology for this population (with a more compact effective radius than elliptical in category VI).  The environmental quenching mechanisms discussed above do not explain the morphological transformation into elliptical. It may in turn favour a rapid morphological transformation by merging for these low-mass galaxies \citep[]{Schawinski2014}. 
\end{itemize}

 As shown in this section, the UV-optical-near-IR colours combined with the morphological information provides a good basis for studying  star formation activity, evolutionary timescales,  and quenching processes of local galaxies. Our description is only qualitative, and we postpone a more detailed analysis of the few hundred galaxies  for which detailed morphological information can be measured to a future work \citep[]{Baillard2011}.

\section{Conclusions}
We presented the multi-wavelength observations collected in the VIPERS region within the W1 and W4 fields of the CFHTLS. The NIR observations were adjusted to provide $K_s$ -band photometry of the VIPERS spectroscopic sources ($i_{AB}\le 22.5$). By comparing them with external deep data, we estimated the depth of our WIRCam to be $K_{s}\sim 22$ at 3$\sigma$. We combined this dataset with GALEX observations. Because of the large GALEX PSF, we developed a specific procedure to derive GALEX photometry in crowded fields based on $u$ -band optical priors.  We estimated the reliability of the method with simulations and found that we reach robust flux measurement down to  $NUV\sim 25$ at $5\sigma$. With this multi-wavelength catalogue (GALEX, optical, $K_s$) we estimated new reliable photometric redshifts for more than one million galaxies at $z\le 1.5$, with $\sigma_z \le 0.04$ and $\eta \le 2\%$ down to $i \sim 23$.
We also pushed our analysis at higher redshift with the BzK galaxy populations.  Despite our native $gri$ detection method, we showed that we can assemble a complete sample of BzK star-forming galaxies  down to $K_s\sim22$ and BzK passive galaxies down to $K_s\sim 20.7$,

 The deep multi-wavelength dataset combined with the unprecedented large area makes this sample perfectly suited for statistical analysis and constrains the rare massive galaxies.  In the companion paper \citep{Moutard2016b}, we use this sample to determine the massive end of the GSMF for quiescent and star-forming galaxies based on photometric redshifts at $0.2\le z\le 1.5  $ with high accuracy, while \citet{Davidzon2013} performed a similar analysis in between $0.5\le z\le 1.3$ with the VIPERS spectroscopic sample. 
The main feature of the VIPERS Multi-Lambda Survey is its ability 
 to investigate the galaxy properties as a function of environment at high redshift. 
 \citet[]{Davidzon2016} use the VIPERS spectroscopic redshifts to investigate the changes in the GSMFs in high and low density regions at $z\sim 1$ while \citet[]{Coupon2015} use photometric redshifts combined with galaxy clustering and galaxy-galaxy lensing  to provide new constraints on the relation between galaxies and their host dark matter halos. In a forthcoming work, we will reconstruct the cosmic web from the complete VIPERS spectroscopic sample and investigate how galaxy properties change in dependance on their location within the cosmic web. 

 We released the photometry of the  VIPERS Multi Lambda Survey over its 24 deg$^2$. 
 The images, catalogues, and photometric redshifts for $\sim 1.5$ million [($NUV \le 25$) $\cup$ ($K \le 22$)] sources are available at this URL: \url{http://cesam.lam.fr/vipers-mls/}.


\begin{acknowledgements}
We gratefully thank C. Moreau from CeSAM/LAM for her major contribution to the building of the database.
We would like to thank M. Jarvis and B. Haeussler for giving us access to the VIDEO dataset in the XMMLSS field.
We also wish to thank R. Overzier for providing us semi-analytical simulations, and L. Arcila-Osejo for helping us with the BzK compilations. 
This research is in part supported by the Centre National d'Etudes Spatiales (CNES) and the Centre National de la Recherche Scientifique (CNRS) of France, and the ANR Spin(e) project (ANR-13-BS05-0005, http://cosmicorigin.org).
L.G. acknowledges support of the European Research Council through the Darklight ERC Advanced Research Grant (\# 291521).
This paper is based on observations obtained with MegaPrime/MegaCam, a joint project of CFHT and CEA/DAPNIA, and with WIRCam, a joint project of CFHT, Taiwan, Korea, Canada and France, at the Canada-France-Hawaii Telescope (CFHT).The CFHT is operated by the National Research Council (NRC) of Canada, the Institut National des Science de l'Univers of the Centre National de la Recherche Scientifique (CNRS) of France, and the University of Hawaii. 
This work is based in part on data products produced at TERAPIX available at the Canadian Astronomy Data Centre as part of the Canada-France-Hawaii Telescope Legacy Survey, a collaborative project of NRC and CNRS. 
We thank the TERAPIX team for the reduction of all the WIRCAM images and the preparation of the catalogues matching with the T0007 CFHTLS data release.
This paper is based on observations made with the Galaxy Evolution Explorer (GALEX). GALEX is a NASA Small Explorer, whose mission was developed in cooperation with the Centre National d'Etudes Spatiales (CNES) of France and the Korean Ministry of Science and Technology. GALEX is operated for NASA by the California Institute of Technology under NASA contract NAS5-98034.
This paper uses data from the VIMOS Public Extragalactic Redshift Survey (VIPERS). VIPERS has been performed using the ESO Very Large Telescope, under the "Large Programme" 182.A-0886. The participating institutions and funding agencies are listed at http://vipers.inaf.it.
This paper uses data from the VIMOS VLT Deep Survey (VVDS) obtained at the ESO Very Large Telescope under programs 070.A-9007 and 177.A-0837, and made available at the CeSAM data center,  Laboratoire d'Astrophysique de Marseille, France.
Funding for PRIMUS is provided by NSF (AST-0607701, AST-0908246, AST-0908442, AST-0908354) and NASA (Spitzer-1356708, 08-ADP08-0019, NNX09AC95G). Funding for SDSS-III has been provided by the Alfred P. Sloan Foundation, the Participating Institutions, the National Science Foundation, and the U.S. Department of Energy Office of Science. The Participating Institutions of the SDSS-III Collaboration are listed at http://www.sdss3.org/.
\end{acknowledgements}

\bibliography{bib_VMLS_data} 
%

\end{document}